\documentclass[aps,prb,twocolumn,reprint,showpacs,superscriptaddress,dvipsnames]{revtex4-2}

\usepackage{epsfig}
\usepackage{amsmath,amssymb}
\usepackage{graphicx}
\usepackage{hyperref}
\hypersetup{colorlinks=true,urlcolor=blue,citecolor=blue,linkcolor=blue,breaklinks=true}
\usepackage[noabbrev]{cleveref}
\usepackage{footnote}
\usepackage{grffile}
\usepackage{color}
\usepackage{dsfont}
\usepackage{mathtools}
\usepackage{verbatim}
\usepackage{bbold}
\usepackage{hhline}
\usepackage{textcomp}
\usepackage{braket}
\usepackage{appendix}
\usepackage[normalem]{ulem}
\usepackage{xcolor}

\allowdisplaybreaks

\begin{document}

%--------
\title{Kohn-Luttinger-like mechanism for unconventional charge density waves 
}
%--------

\newcommand{\TUM}{\affiliation{School of Natural Sciences, Technische Universit\"at München, D-85748 Garching, Germany}}

\newcommand{\MPI}{\affiliation{Max Planck Institute for Solid State Research, D-70569 Stuttgart, Germany}}

\newcommand{\RUB}{\affiliation{Institute for Theoretical Physics III, Ruhr-University Bochum, D-44801 Bochum, Germany}}

\author{Hannes Braun}       \MPI\TUM
\author{Michael M. Scherer} \RUB
\author{Laura Classen}       \MPI\TUM

%--------

\begin{abstract} Interaction-induced charge orders with electronic origin occur as states of spontaneously broken symmetry in several materials platforms. An electronic mechanism for charge order requires an attractive component in the effective charge vertex. We put forward such a mechanism for the formation of unconventional charge density waves in a metal. These states result from the condensation of particle-hole pairs  with finite wave vector and nonzero angular momentum and correspond to bond or loop current order on a lattice. The mechanism we describe can be viewed as Kohn Luttinger analysis in the particle-hole channel with finite transferred momentum. It incorporates one-loop spin and pairing corrections, which are then used as an input for a summation in the charge channel triggering an instability. We extend our analysis to a spin-fluctuation approach, where the effective charge interaction is dressed by the particle-hole ladder with exchanged momentum. We argue that this mechanism works for weakly interacting metals with a nested Fermi surface and a large number of fermion flavors. We apply the Kohn-Luttinger-like approach to square- and triangular-lattice Hubbard models with SU($N_f$) flavour symmetry and show that it leads to different types of $p$-wave charge density waves. We also study effects beyond weak coupling at and away from the Van Hove filling in terms of a phenomenological model with additional exchange interaction. In the vicinity of Van Hove filling, we obtain $d$-wave charge density waves with wave vectors determined by nesting as leading instabilities. In addition, we find another charge density wave with wave vector $K/4$ on the triangular lattice on both sides of Van Hove filling. We demonstrate that this $K/4$  instability can win the competition against pairing for $N_f=4$ via an unbiased functional renormalization group calculation. \end{abstract}

\maketitle

%--------
\section{Introduction}
%--------

Fascinating states of matter can arise from spontaneous symmetry breaking at low temperatures in an interacting metal. 
They occur in different forms such as spin, charge, or pairing orders.  Studying the mechanisms behind their formation provides insights about the fundamental processes in a quantum  many-body system. 
A famous example is superconductivity from repulsive interactions, which relies on the renormalization of an initially repulsive electron-electron-interaction into an effective attraction between electrons with nonzero relative angular momentum~\cite{Maiti2013,RevModPhys.84.1383,Kohn1965,PhysRev.150.202}.
In contrast, spin orders can directly profit from the bare Coulomb interaction in the sense that an instability is, in principle, possible 
if the interaction is large enough. 
The Stoner criterion is a well-known example for this. 
For charge orders it is less common to become the leading instability. 
One reason is that the standard screening within a random phase approximation (RPA) does not lead to an instability for a repulsive Coulomb interaction. 
However, Pomeranchuk instabilities with higher angular momentum that distort the Fermi surface or charge density waves~(CDW) can occur due to additional renormalization effects~\cite{Pomeranchuk1958,Chubukov2018,PhysRevB.74.115126}.

Here, we are interested in a mechanism that can lead to a charge-ordered ground state, in particular, one that breaks translation symmetry. 
Such a charge density wave results from the condensation of particle-hole pairs with a finite wave vector. 
Interestingly, the particle-hole pairs can also possess a non-zero relative angular momentum~\cite{PhysRevB.62.4880}. 
We call the corresponding states unconventional charge density waves (uCDW) in analogy to the case of superconductivity from repulsive interactions. 
They need to be distinguished from charge Pomeranchuk orders~\cite{Pomeranchuk1958,Chubukov2018}, which are also based on particle-hole pairs with different angular momenta but have zero wave vector, i.e., they do not break translation symmetry. 
Consequently, a symmetry classification differs between Pomeranchuk and CDW orders~\cite{Venderbos2016}. 
Examples for uCDWs include bond or current orders, and were investigated in many systems. 
For example, a Kekul\'e valence bond state can occur on the honeycomb lattice and was discussed for graphene~\cite{PhysRevLett.98.186809,PhysRevB.80.205319,PhysRevB.81.085105,PhysRevB.88.245123,PhysRevLett.107.106402,PhysRevB.90.035122,PhysRevLett.111.066401}. 
More recently, an incommensurate more complex version, the incommensurate Kekul\'e spiral \cite{PhysRevX.11.041063} was observed in twisted bilayer graphene \cite{Nuckolls2023}.  
Current orders spontaneously break time-reversal symmetry and they induce flux patterns if they form loop currents.
Loop current orders or staggered flux states, as well as bond orders, were discussed in the context of the cuprates \cite{PhysRevB.55.14554,PhysRevB.67.054511,PhysRevB.73.155113,PhysRevB.81.064515,PhysRevB.75.224511,PhysRevLett.96.197001,PhysRevB.78.020506,LiGreven2008,LiGreven2010,CRPHYS_2021__22_S5_7_0,PhysRevB.96.214504,PhysRevLett.111.027202,Palle2024}. 
Furthermore, their topological properties were studied and it was pointed out that topological density waves provide as a way to obtain interaction-induced topological states 
\cite{PhysRevLett.100.156401,PhysRevB.82.195126,PhysRevLett.107.106402,PhysRevB.84.155111,LiSarma2015,Lin2019,Classen2019}. 
Recently, charge order that breaks translation and maybe also time-reversal symmetry was observed in kagome metals \cite{neupert2022,PhysRevMaterials.3.094407,PhysRevLett.125.247002,YangMazhar2020,JiangZahid2021,PhysRevX.11.031050,PhysRevX.11.031026,PhysRevX.11.041030,PhysRevB.104.L041103,LiMiao2022,Mielke2022,PhysRevResearch.4.023244,PhysRevLett.131.016901,Xing2024,Gupta2022,Kundu2024}, where an interesting interplay of multi-Q CDWs can occur \cite{PhysRevB.104.214513,PhysRevB.104.045122,PhysRevB.106.144504}.  

A way to understand the formation of uCDWs out of a metallic state is to consider them as interaction-induced instabilities of a Fermi liquid. 
In this scenario, the effective interaction in the charge channel needs to develop a singularity at a finite wave vector and higher angular momentum. 
However, an RPA summation does not result in such a singularity for a repulsive density-density interaction with a (screened) Coulomb-like profile. 
This indicates that the feedback from spin and charge channels plays an important role for the formation of uCDWs. 
An analogous mechanism can induce nematic order \cite{PhysRevB.98.041108,grandi2024} and assist in pairing near a nematic instability \cite{islam2023}.
Inter-channel feedback is taken into account in parquet or functional renormalization group (PRG/FRG) studies, which indeed reported charge bond order in the phase diagram on the kagome lattice~\cite{PhysRevLett.110.126405,PhysRevB.87.115135,profe2024}, or current loop orders on the square and triangular lattice~\cite{Honerkamp2004p1,Lin2019,Classen2019}.

For strongly interacting systems, another route towards uCDWs is given via an exchange interaction between neighboring spins, which can also induce attractive components in the charge channel. 
This was analysed on the square lattice with respect to Peierls or staggered flux states (a.k.a. bond and current orders) ~\cite{PhysRevB.37.3774,PhysRevB.39.11538,PhysRevB.43.10436,PhysRevB.43.2866,PhysRevB.59.6475,PhysRevLett.90.186401,PhysRevB.68.132501}. 
In the examples above, uCDWs compete with pairing or spin-density-wave (SDW) instabilities. 
Thus, for them to win the competition the effective coupling for uCDWs needs to be larger than for pairing or SDW. 
This can be achieved, for example, via an increased number of fermion flavours which boost the uCDW coupling~\cite{Honerkamp2004p1,Lin2019,Classen2019}, or via multi-orbital/sublattice interference effects which suppress competing SDW and pairing couplings~\cite{PhysRevB.86.121105}.
It is also possible to engineer such conditions, e.g, in correlated fermion systems with SU($N_f$) flavour symmetry based on ultra-cold atoms,  which can realize $N\leq 10$ \cite{doi:10.1126/science.1254978,coldatomsSUN,PhysRevX.6.021030,coldatomsSUN1D} and SU($2$)$\times$SU($6$) mixtures~\cite{taie2010realization,Cazalilla_2014}, or moir\'e heterostructures,
 which can approximately realize SU($4$) symmetry~\cite{Zhang2021}. In rare examples, SU($N_f$) with $N_f>2$ also occur
in solid state systems \cite{PhysRevB.90.094422}, such as $\alpha$-ZrCl${}_3$ \cite{PhysRevLett.121.097201}. 
Theoretical analyses of Fermi liquid instabilities in SU($N_f$) systems were performed using multiple methods \cite{Honerkamp2004p1,Classen2019,Lin2019}.

In this report, we present yet another mechanism to obtain uCDWs from repulsive interactions inspired by the Kohn-Luttinger (KL) mechanism for superconductivity~\cite{Kohn1965,PhysRev.150.202}. 
We start with a weakly interacting metallic system and discuss under which conditions an instability towards uCDW order can arise. 
To this end, we assume that there is a large number of fermion flavours $N_f$ and that the single-particle dispersion is at least approximately nested. 
We argue that in this situation an attraction in the uCDW channel can already occur when spin and pairing corrections are included on the one-loop level and determine the instability condition for a general bare interaction. 
As a next step, we consider higher-loop corrections in terms of a spin-fluctuation approach, i.e., we use the effective interaction in the spin channel as input for the charge channel. 

Our analysis works out the minimal fluctuations that are required for an uCDW instability and make the delicate interplay between spin and pairing corrections clearly visible. 
In this sense, they complement PRG and FRG approaches which take into account all channels on equal footing. 
We apply this KL-like approach to the square- and triangular-lattice Hubbard model, where we find that it results in $p$-wave CDW with wave vector given by the nesting vectors. 
On the square lattice, $p_x$ and $p_y$ solutions are degenerate and describe bond or current density waves depending on their superposition.
On the triangular lattice, there is a unique $p$-wave form factor for each of the three non-equivalent nesting vectors. 
They separately form bond density waves, while the exact ground state also depends on their superposition. 

Finally, we complement our KL-like analysis by studying a $t-U-J$ model with Hubbard $U$ and SU($N_f$) nearest-neighbor exchange interaction as a phenomenological description beyond weak coupling. 
For the square lattice, we reproduce earlier results for the staggered flux state \cite{PhysRevB.37.3774,PhysRevB.39.11538}. 
For the triangular lattice, we find a phase transition between uCDW with different wave vectors as a function of the chemical potential. 
The first one occurs at Van Hove filling and corresponds to the flux order with nesting wave vectors $M_i$ that was discussed previously \cite{Venderbos2016,Lin2019,Classen2019}. 
Moving away from the Van Hove filling, the maximum of the susceptibility jumps from $M_i$ to $K/4$, which is why another uCDW with wave vector $K/4$ and $d$- or $f$-wave form factor becomes the leading instability. 
We confirm that the $d$-wave $K/4$ CDW can win against pairing in an FRG calculation for the SU(4) case.

In the following, we first introduce the general formalism and describe the mechanisms for uCDW instabilities in Sec.~\ref{sec:formalism}. 
We also compare to the original KL mechanism for pairing instabilities. 
In Sec.~\ref{sec:applylattice}, we present the application to the square and triangular lattice model, as well as the analysis of the phenomenological $t-U-J$ model. 
We summarize our findings and comment on extensions of our approach in Sec.~\ref{sec:conclusion}. 
In the appendix, we present a KL-like analysis for uCDWs in a patch model for square and triangular lattice and compare to the results of full model in the main text. 

%--------
\section{Formalism and idea} \label{sec:formalism}
%--------

%--------
\subsection{Effective interaction and form-factor expansion}
%--------

%
\begin{figure*}
    \centering
\includegraphics[width=1\textwidth]{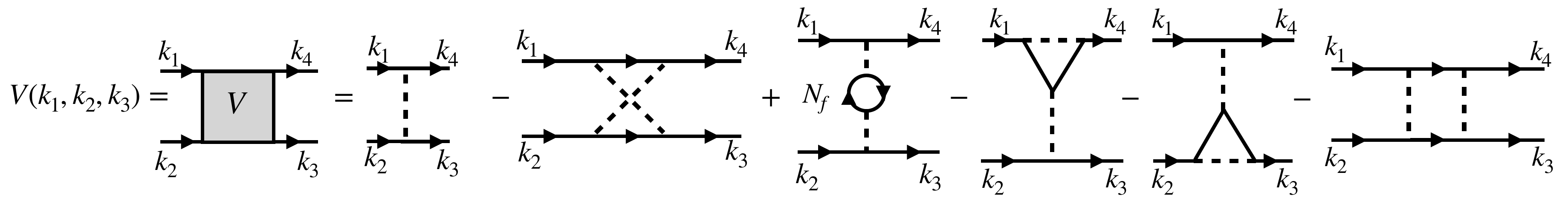}
    \caption{\textbf{Effective interaction} $V(k_1,k_2,k_3)$ expanded up to one-loop order. The first term on the left-hand side depicts the contribution from the bare interaction $U(k_1,k_2,k_3)$ (dashed line)and all other terms are one-loop corrections. External solid lines determine the in-going and out-going momenta of the interacting fermions, 
    and internal solid lines in the loops represent single-particle Green's functions $\mathcal{G}$. Notably, the closed internal loop in the first direct particle-hole diagram, which is also known as ``screening'' diagram, contributes with the multiplicity of the flavour number $N_f$.}
    \label{fig:oneloopcorrections}
\end{figure*} 

In a many-body system, interactions receive corrections from multiple scattering processes. Formally, this can be described by the vertex function $\Gamma^{(2)}_{\alpha\beta\gamma\delta}(k_1, \dots ,k_4)$ which is the one-particle irreducible part of the two-particle Green's function
\begin{align}
G^{(2)}_{\alpha\beta\gamma\delta}(k_1, \dots ,{k}_4)=\langle \hat T c_{{k}_1 \alpha}^\dagger c_{{k}_2 \beta}^\dagger c_{{k}_3 \gamma} c_{{k}_4 \delta} \rangle\,,
\end{align}
where $\hat T$ is the time-ordering operator and the $c^{(\dagger)}_{\alpha k}$ are fermion annihilation (creation) operators with spin $\alpha$ and (2+1)d-momentum $k=(\omega,\mathbf k)$ collecting Matsubara frequency and wave vector. 
Flavour symmetry SU($N_f$), such as spin SU(2), dictates the form 
\begin{align}
\Gamma_{\alpha\beta\gamma\delta}(k_1, \dots ,k_4)&=V(k_1,k_2,k_3,k_4)\delta_{\alpha\delta}\delta_{\beta\gamma}\notag\\
&\quad-V(k_1,k_2,k_4,k_3)\delta_{\alpha\gamma}\delta_{\beta\delta}\,,
\end{align}
and we can view $V(k_1,k_2,k_3,k_4)$ as the effective interaction.  It appears in the effective action as
\begin{align}
    \Gamma_V &= \frac{1}{2}\int_{{k}_1, \dots ,{k}_4} \sum_{\sigma,\sigma'} V({k}_1, \dots ,{k}_4) \delta({k}_1+{k}_2 - {k}_3-{k}_4) \notag\\
    &\quad\times c_{{k}_1 \sigma}^\dagger
    c_{{k}_2 \sigma'}^\dagger
    c_{{k}_3 \sigma'}
    c_{{k}_4 \sigma}\,, \label{eq:action}
\end{align}
and we use the short-hand notation $T\sum_\omega \frac{1}{V}\int d\mathbf{k}\equiv \int_{k}$. 

Due to momentum conservation, the interaction depends effectively only on three momenta, which we use to shorten notation by omitting the fourth one in the following.  
To first order, the effective interaction equals the bare interaction $V(k_1,k_2,k_3)=U(k_1,k_2,k_3)$. 
Below we will consider a constant Hubbard interaction $U(k_1,k_2,k_3)=U$ and study the repulsive case $U>0$. To second order, $V(k_1,k_2,k_3)$ receives the one-loop corrections shown diagrammatically in Fig.~\ref{fig:oneloopcorrections}.
They can be decomposed into contributions of particle-particle type
\begin{align}
    &\tau_{\mathrm{pp}}(k_1,k_2,k_3)=-\int_{p}
    \mathcal{G}({p}+Q_{\mathrm{pp}})\mathcal{G}(-{p}) \notag\\
    &\quad\quad\quad\quad\times U({k}_1 , {k}_2, Q_{\mathrm{pp}}-{p}) U({p},Q_{\mathrm{pp}}-{p},  {k}_3)\, , \label{eq:tpp}
\end{align}
of crossed particle-hole type
\begin{align}
    &\tau_{\mathrm{ph-c}}(k_1,k_2,k_3)=-\int_p \mathcal{G}({p}+Q_{\mathrm{ph-c}})\mathcal{G}({p}) \nonumber\\
    &\quad\quad\quad\quad\times U({k}_1,{p},k_3) U({p}+Q_{\mathrm{ph-c}},{k}_2,p)\,,\label{eq:contribution2}
\end{align}
and of direct particle-hole type
\begin{align}
    &\tau_{\mathrm{ph-d}}(k_1,k_2,k_3) =-\int_p \mathcal{G}( {p}+Q_{\mathrm{ph-d}})\mathcal{G}({p}) \nonumber\\
     &\quad\quad\quad\quad \times\bigl[U({k}_1,{p}+Q_{\mathrm{ph-d}},p) U(k_2,p,k_3) \nonumber\\
&\quad\quad\quad\quad+ U({k}_1,{p}+Q_{\mathrm{ph-d}},k_4) U(p,k_2,k_3)\nonumber\\
&\quad\quad\quad\quad-N_f U({k}_1, {p}+ Q_{\mathrm{ph-d}}, {p}) U({p},{k}_2,{k}_3) \bigr]\,.\label{eq:tphd}
\end{align}
Here, $\mathcal{G}(k)=(i\omega-\xi_\mathbf{k})^{-1}$ is the bare
%HB do we need to write G_0?
single-particle Green's function with dispersion $\xi_\mathbf{k}$
and we introduced the total, transferred, and exchanged momentum,  $Q_{\mathrm{pp}}=k_1+k_2$, $Q_{\mathrm{ph-c}}={k}_1-{k}_3$, and $Q_{\mathrm{ph-d}}={k}_2-{k}_3$, respectively. 

Generally, the effective interaction can be viewed as a function of either total, transferred, or exchanged momentum according to these corrections. This defines three channels $V_X$, $X \in \{ \text{pp, ph-d, ph-c}\}$
\begin{align}
    V_{\text{pp}}(k,k',Q)&=V({k} , {Q} -{k},{Q} -{k}', {k}')  \\
        V_{\text{ph-c}}(k,k',Q)&=V({k}, {k}'+Q, {k}+ {Q}, {k}') \\
        V_{\text{ph-d}}(k,k',Q)&=V({k}, {k}'+Q, {k}', {k}+{Q})\,, 
\end{align}
where the effective interaction $V_X$ is two-particle reducible in the particle-particle (pp), the direct particle-hole (ph-d), and the crossed particle-hole (ph-c) channel, respectively. Note that in a single-channel approximation, the effective interaction can be obtained from a differential equation, where the change of $V$ with a scale $T$ is expressed via $\partial_T V_X=\tau_{X}^T$. This amounts to a resummation of an infinite series in the bare interaction $U$. In an FRG approach, the differential equation contains all three channels on equal footing [see Sec.~\ref{sec:tju}, Eq.~\eqref{eq:FRG} below].

This rewriting allows us to single out the decisive momentum ${Q}$ and two less important momenta ${k}, {k}'$ for an instability in each channel. At low temperatures, one or more of the $V_X$ can develop a singularity at a specific $Q$, which signals the tendency to form an ordered phase with particle-hole or particle-particle bound states. To study the most singular correction, we set external frequencies to the lowest Matsubara frequency as the frequency dependence of the vertex functions 
has a subleading effect on the primary divergence structure in weak coupling and we focus on the dominant contribution near the critical temperature. The results of such an approximation were shown to yield qualitatively robust results upon inclusion of frequencies at weak coupling \cite{PhysRevB.85.075121,PhysRevResearch.2.033372,10.21468/SciPostPhys.6.1.009,PhysRevB.102.245128}.
However, we note that the frequency dependence becomes important for stronger coupling and can qualitatively affect the nature of correlations \cite{PhysRevLett.110.246405,PhysRevB.95.235109,PhysRevLett.119.056402}.

For example, the Cooper instability corresponds to a singularity in $V_{\mathrm{pp}}$ for $Q=0$. 
The particle-hole channels are related to instabilities towards charge and spin or more generally flavour order. This can be seen when rewriting the effective action in Eq.~\ref{eq:action} via charge and flavour components
\begin{align}
    \Gamma_V &= \frac{1}{2}\int_{k, k' ,q} \sum_{\alpha\ldots\delta} c_{{k} \alpha}^\dagger
    c_{{k}'+q \beta}^\dagger
    c_{k' \gamma}
    c_{k+q \delta} \notag\\ 
    &\left[V^c(k, k' ,q)\delta_{\alpha\delta}\delta_{\beta\gamma} - V^s(k, k' ,q)\sum_{i=1}^{N_f^2-1}T^i_{\alpha\delta}T^i_{\beta\gamma}  \right]\,,\label{eq:chargespinvertex}
\end{align}
where $T^i$ are the generators of SU($N_f$) in the defining representation and we used the completeness relation $2T^i_{\alpha\beta}T^i_{\gamma\delta}+\delta_{\alpha\beta}\delta_{\gamma\delta}/N_f=\delta_{\alpha\delta}\delta_{\beta\gamma}$ \cite{SUNidentities}. For SU(2), these are simply proportional to the Pauli matrices $T^i=\sigma^i/2$. The effective interactions in charge and flavour channel are given by $V^c(k, k' ,q)=[V_{\mathrm{ph-d}}(k, k' ,q)-V_{\mathrm{ph-c}}(k, k' ,q)/N_f]/2$ and $V^s(k, k' ,q)=-V_{\mathrm{ph-c}}(k, k' ,q)$. This means that the effective interaction in the charge channel is well approximated by the effective interaction in the ph-d channel for large $N_f$.

\begin{figure*}
    \centering
    \includegraphics[width=0.9\linewidth]{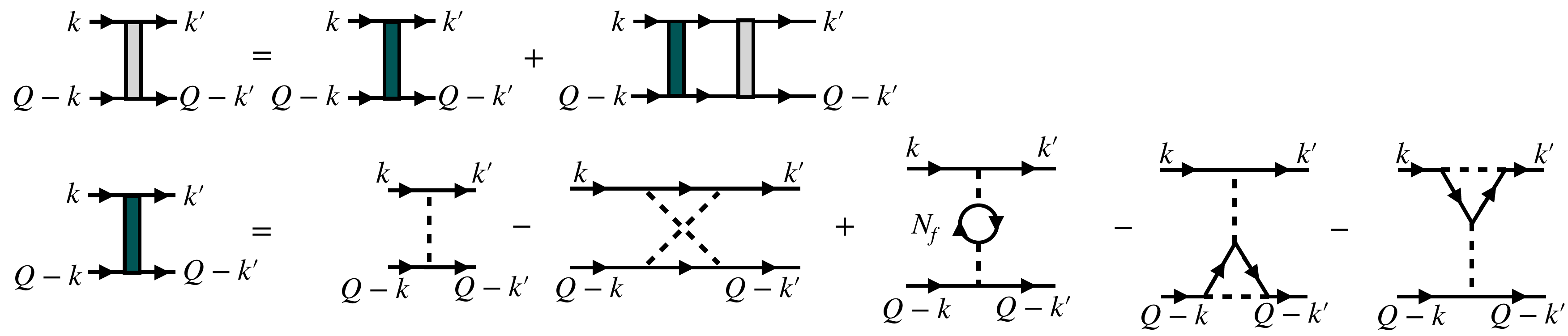}
    \caption{\textbf{Kohn-Luttinger mechanism for pairing.} Top: diagrammatic representation of the self-consistent equation for the effective pairing vertex $V_{pp}$, i.e., the effective interaction viewed as a function of total incoming/outgoing momentum. The self-consistent equation amounts to summing up the Cooper ladder. There is no pairing instability with only a bare Coulomb interaction (dashed line). Using instead the bare interaction and one-loop particle-hole corrections (bottom) as a basis in the Cooper ladder (blue vertex), a pairing instability for pairs with non-zero angular momentum can occur. Internal solid lines represent fermion propagators, external ones denote the momentum dependence.}
    \label{fig:KLpairing}
\end{figure*}

It is convenient to expand the momentum dependence of $V_X(k,k',Q)$ in form factors according to the symmetry of the system. For example, we choose the lattice harmonics of the square and triangular lattice in Sec.~\ref{sec:applylattice} below. This can be thought of as an expansion in the angular momentum generalized to a lattice system. 
For the expansion, we project the vertex into form factor space
\begin{align}
      \hat{V}_X^{l,l'}(\mathbf{Q})=\int_{\mathbf{k}, \mathbf{k}'}f_l(\mathbf{k})f^\ast_{l'}(\mathbf{k}')V_X(\mathbf{k, \mathbf{k}',\mathbf{Q})} \label{eq:ffexpansion}
\end{align}
where the form factors $f_i(\mathbf{k})$ form an orthonormal basis $\delta(\mathbf{k}-\mathbf{k}') = \sum_l f_l^\ast ( \mathbf{k}) f_l(\mathbf{k}') $ and 
$\delta_{l,l'} =\int_\mathbf{k} f_l^\ast(\mathbf{k}) f_{l'}(\mathbf{k})$. 
This treatment is inspired by the truncated unity approach to the fRG \cite{Lichtenstein2017}, but can be done independently and is helpful in order to get a better understanding of the symmetries of the instabilities.
Analogously, we project the particle-particle (-) and particle-hole (+) polarizations into the form-factor basis for later use
\begin{equation}
    \hat{\Pi}_\pm^{l,l'}(\mathbf{Q})= \mp\int_{{p}}\mathcal{G}({p})\mathcal{G}(\pm{p}+Q)f^\ast_l(\mathbf{p})f_{l'}(\mathbf{p}) \label{eq:bubble_in_ff}
\end{equation}
To a good approximation, it is usually sufficient to consider a finite number of form factors.
Here, we have chosen the signs of the particle-particle and particle-hole polarisations such that $\Pi_\pm(\mathbf{Q})>0$.

%--------
\subsection{Kohn Luttinger mechanism for pairing}
%--------

Even in a weakly interacting system of fermions with purely repulsive, bare interactions, instabilities of the Fermi liquid arise at low temperatures. The reason is that the effective interaction becomes attractive in a pairing channel due to corrections from particle-hole channels, which then gives rise to a Cooper instability. 
Kohn and Luttinger showed that the decisive corrections come from a non-analytic contribution in the particle-hole polarization at a momentum transfer of twice the Fermi momentum $q=2k_F$, which possesses attractive components in higher angular-momentum channels. 
To see this, we consider the pairing channel, cf.~Fig.~\ref{fig:KLpairing},
\begin{align}
    &V_{\text{pp}}(\mathbf{k} , \mathbf{k}', \mathbf{Q}) = U(\mathbf{k} , \mathbf{Q}-\mathbf{k}, \mathbf{Q}-\mathbf{k}')\notag \\
    &-\!\int_{p} U(\mathbf{k} , \mathbf{Q}\!-\!\mathbf{k},\mathbf{Q}\!-\!\mathbf{p})\mathcal{G}({p})\mathcal{G}(Q\!-\!{p}) V_{\text{pp}}(\mathbf{p},  \mathbf{k}', \mathbf{Q})\,, \label{eq:ppladder}
\end{align}
and project it into the form-factor basis
\begin{align}
    \hat{V}^{l,l'}_{\text{pp}}(\mathbf{Q}) = \hat{U}_{\mathrm{pp}}^{l,l'}(\mathbf{Q}) - \hat{U}_{\mathrm{pp}}^{l,a}(\mathbf{Q}) \hat{\Pi}_-^{a,b}(\mathbf{Q}) \hat{V}_{\text{pp}}^{b,l'}(\mathbf{Q})\,,\label{eq:pp_in_ff}
\end{align}
where summation over repeated indices is implied and the bare interaction in the particle-particle channel reads
\begin{align}
\hat{U}_{\mathrm{pp}}^{l,l'}(\mathbf{Q})=\int_{\mathbf{k,'k}}f_l(\mathbf{k})f^\ast_{l'}(\mathbf{k}')U(\mathbf{k},\mathbf{Q}-\mathbf{k},\mathbf{Q}-\mathbf{k}')\,.
\end{align}
As long as the interactions consist only of static, bare Coulomb interactions and bare propagators can be used, the right hand side of Eq. \ref{eq:pp_in_ff} remains independent of external frequencies.
The formal solution in matrix representation is given by
\begin{align}
    \hat{V}_{\text{pp}}(\mathbf{Q}) 
    &=\frac{\hat{U}_{\mathrm{pp}}(\mathbf{Q}) }{\mathbb{1}+  \hat{U}_{\mathrm{pp}}(\mathbf{Q}) \hat{\Pi}_-(\mathbf{Q}) }\,. \label{eq:ppladder_frac}
\end{align}
We note that for an isotropic system and $\bf{Q}=0$, as considered by KL, all quantities become diagonal when expanding in angular harmonics, so that the $l$th angular-momentum component reads $\hat V^l_{\mathrm{pp}}=\hat U_{\mathrm{pp}}^l(1+\hat U_{\mathrm{pp}}^l \hat \Pi_l)^{-1}$. 

The bare Coulomb repulsion does not possess an attractive angular-momentum component and with it alone no pairing instability can arise.  
However, taking one-loop particle-hole corrections into account, we have to replace the bare interaction in the pairing channel by
\begin{align}
U_{\mathrm{pp}} \rightarrow  V_{\mathrm{pp}}^{0}=  U_{\mathrm{pp}} + \left.\tau_{\mathrm{ph-c}}\right|_{\mathrm{pp}} + \left.\tau_{\mathrm{ph-d}}\right|_{\mathrm{pp}}\,,
\end{align}
where $\tau_{X}|_{\mathrm{pp}}=\tau_{X}(k,Q-k,Q-k')$, see Fig.~\ref{fig:KLpairing}. Explicitly, this reads
\begin{align}
    \hat V_{\text{pp}}^{0,l,l'}(\mathbf{Q})&=U\delta_{l,0}\delta_{l',0} +U^2\int_{\mathbf{k},\mathbf{k}'}f_l(\mathbf{k})f^\ast_{l'}(\mathbf{k}')\notag\\
    &\quad \times\left[\Pi_+(\mathbf{k}+\mathbf{k}')+(N_f-2)\cdot\Pi_+(\mathbf{k}-\mathbf{k}')\right]\ \label{eq:Veffllpp}.
\end{align}
where the term with prefactor $N_f-2$ belongs to the ph-d fluctuations. The other $\mathcal{O}(U^2)$ term corresponds to the spin fluctuations.
This is not only the point where the effective interaction becomes momentum dependent, but it would also develop a non-trivial frequency dependence
with a form analogous to the momentum dependence.  
As stated above, we limit ourselves to the lowest Matsubara frequencies and neglect self-energy effects. 
This mean that we exclude any odd-frequency pairing.

\begin{figure*}
    \centering  \includegraphics[width=0.9\linewidth]{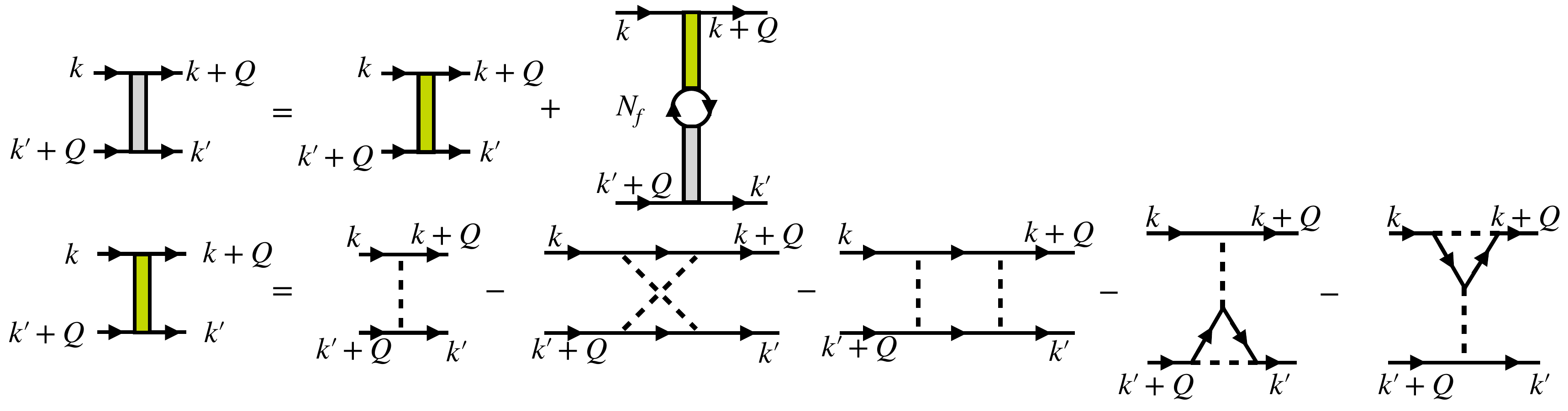}
    \caption{\textbf{Kohn-Luttinger-like mechanism for uCDWs.} Top: diagrammatic representation of the large-$N_f$ self-consistent equation for the effective interaction viewed as a function of transferred momentum, i.e., in the direct particle-hole channel $V_{\mathrm{ph-d}}$. In the large-$N_f$ limit $V_{\mathrm{ph-d}}$ is a good approximation for the effective charge vertex. The self-consistent equation is equivalent to summing the RPA screening diagrams in the direct particle-hole channel. Bottom: similar to the Kohn-Luttinger mechanism for pairing, an instability can arise in the RPA summation if the bare interaction (dashed line) is supplemented by one-loop particle-particle and remaining particle-hole diagrams (green vertex). These diagrams are subleading in $N_f$. Internal solid lines represent fermion propagators, external ones denote the momentum dependence.
    }
    \label{fig:uCDW_ladder}
\end{figure*}

KL showed that in an isotropic Fermi liquid, the bare repulsion $\hat U_{\mathrm{pp}}^l$ is exponentially small for large $l$ and that $\tau_{\mathrm{ph,c}}|_{\mathrm{pp}} + \tau_{\mathrm{ph,d}}|_{\mathrm{pp}}$ possesses attractive angular-momentum components due to the nonanalyticity of the particle-hole polarization bubble at $\Pi_-^l(k\pm k'\approx 2k_F)$ so that $V_{\mathrm{pp}}^{0,l}<0$ for some $l>0$. As a result, a pairing instability arises even in a weakly interacting Fermi liquid based on a purely electronic pairing mechanism. 

Critical temperatures of this original KL mechanism remain very small, i.e., exponentially small in $l^4$, but they can be boosted through peaks in the density of states in a lattice system when a Van Hove singularity is placed near the Fermi level. Closely related in spirit is also the spin-fluctuation mechanism: the one-loop particle-hole corrections are only the first terms of the renormalization in the particle-hole channel and the entire series should be summed if they are large, as is the case close to a magnetic instability.

%--------
\subsection{Mechanisms for unconventional CDW}\label{sec:KLCDW}
%--------

Similar to the case of pairing, the couplings for charge ordering tendencies are repulsive when starting with a (screened) Coulomb interaction.  
However, following the idea of KL, corrections from other channels can introduce an attraction and consequently induce instabilities towards charge order. 
We show that such a KL-like mechanism also works for finite wave-vector transfer as long as higher lattice harmonics are considered, i.e., the mechanism yields instabilities towards ``unconventional'' charge density waves with order parameters of the form $\Delta_Q(\mathbf{k})=f_l(\mathbf{k})\langle c_{k+Q}^\dagger c_k\rangle$ where $l>0$ and $Q\neq 0$. 
Depending on $l$ and $Q$, these orders describe, for example, bond density waves for real $\Delta_Q$ or loop current/staggered flux phases for imaginary $\Delta_Q$. 

For density-wave instabilities to become competitive at weak coupling, we consider a system with (approximate) nesting, so that particle-particle and particle-hole channels are parametrically of the same order. Specifically, this means that they possess a (double-)logarithmic singularity away from (at) Van Hove filling (and above an infrared cutoff scale associated with deviations from perfect nesting). In addition, we consider a system with SU($N_f$) symmetry where $N_f$ is large, which will give charge instabilities a boost over pairing and spin instabilities. 

In this situation, the leading corrections to the effective interaction are based on the  one-loop ``screening'' diagram, cf. Fig.~\ref{fig:uCDW_ladder}, which is of the order of $N_f \ln^2(T)$ (or $N_f\ln(T)$ without Van Hove singularity). 
Summing up the corresponding ladder series shown in Fig.~\ref{fig:uCDW_ladder}, we obtain the self-consistent equation for the effective interaction in the direct particle-hole channel
\begin{align}
     &V_{\text{ph-d}}(\mathbf{k},\mathbf{k}', \mathbf{Q}) =U(\mathbf{k},\mathbf{k}'+\mathbf{Q},\mathbf{k}')\notag \\
     &-\!N_f\!\!\int_\mathbf{p}
    \!U(\mathbf{k},\mathbf{p}\!+\!\mathbf{Q}, \mathbf{p})
    \mathcal{G}({p}) \mathcal{G}({p}\!+\!{Q})
    V_{\text{ph-d}}(\mathbf{p},\mathbf{k}', \mathbf{Q})\,. \label{eq:phdladder}
\end{align}
We rewrite this in terms of the form factor basis
\begin{align}
    \hat{V}_{\text{ph-d}}^{l,l'}(\mathbf{Q}) 
    &= \hat{U}_{\mathrm{ph,d}}^{l,l'}(\mathbf{Q})\notag\\ 
    &\quad- N_f \hat{U}^{l,a}_{\mathrm{ph,d}}(\mathbf{Q}) \hat{\Pi}^{a,b}_+(\mathbf{Q}) \hat{V}_{\text{ph-d}}^{b,l'}(\mathbf{Q})\,, \label{eq:phdladder_in_ff}
\end{align}
with $\hat{U}_{\mathrm{ph,d}}^{l,l'}(\mathbf{Q})\!=\!\int_{\mathbf{k,'k}}U(\mathbf{k},\mathbf{k}',\mathbf{Q})f_l(\mathbf{k})f^\ast_{l'}(\mathbf{k}')$, and obtain the formal solution in matrix representation
\begin{align}
    \hat{V}_{\text{ph-d}}(\mathbf{Q}) = \frac{\hat{U}_{\text{ph-d}}(\mathbf{Q})}{\mathbb{1} + N_f \hat{ U}_{\text{ph-d}}(\mathbf{Q})\hat{\Pi}_+(\mathbf{Q})}\,. \label{eq:phd_ladder_frac}
\end{align}
Again, there cannot be any instability if the bare interaction does not possess an attractive angular momentum component, which is usually not the case. 
However, analogous to the KL scenario for pairing, interaction corrections from other channels will introduce a nontrivial momentum dependence into the renormalized interaction.
Only now we need to project the crossed particle-hole and particle-particle corrections into the direct particle-hole momentum structure. 

As a first step, we include the corresponding one-loop corrections, see Fig.~\ref{fig:uCDW_ladder}. Then, we obtain that the bare interaction in the direct channel is replaced by
\begin{align}
U_{\mathrm{ph-d}} \rightarrow  V_{\mathrm{ph-d}}^{0}=  U^{(2)}_{\mathrm{ph-d}} \!+\! \left.\tau_{\mathrm{ph-c}}\right|_{\mathrm{ph-d}} \!+\! \left.\tau_{\mathrm{pp}}\right|_{\mathrm{ph-d}}\,,
\end{align}
where $\tau_{X}|_{\mathrm{ph-d}}=\tau_{X}(k,k'+Q,k')$ and in $U^{(2)}_{\mathrm{ph-d}}$ we incorporated the ph-d corrections that are of subleading order $\mathcal{O}(1)$ in $N_f$ and not contained in Eq.~\eqref{eq:phdladder}, i.e., $U^{(2)}_{\mathrm{ph-d}}=U_{\mathrm{ph-d}}-\int_p \mathcal{G}(p)\mathcal{G}({p}+Q)[U({k},{p}+Q,p) U(k'+Q,p,k') + U({k},{p}+Q,k+Q) U(p,k'+Q,k')]$, cf.~Eq.~\eqref{eq:tphd} or the last two diagrams in Fig.~\ref{fig:uCDW_ladder}. 
In the case of a pure Hubbard interaction $U(k_1,k_2,k_3)=U$, we find 
\begin{align}
    &V_{\text{ph-d}}^0(\mathbf{k},\mathbf{k}',\mathbf{Q})=U+2U^2\Pi_+(\mathbf{Q}) \notag \\
    &\quad\quad\quad\quad+ U^2[\Pi_+(\mathbf{k}-\mathbf{k}')-\Pi_-(\mathbf{k}+\mathbf{k}'+\mathbf{Q})]\,.
\end{align}
In this case, only the ph-c and pp corrections lead to an additional momentum dependence, which can induce an attraction in a higher harmonic contribution. 
The remaining ph-d corrections lead to a constant shift $2U^2\Pi_+(Q)$, which changes the $s$-wave part of $V^0_{\text{ph-d}}$ to more repulsive.
Explicitly, we find after projecting into form factors
\begin{align}
    &\!\!\hat V_{\text{ph-d}}^{0,l,l'}(\mathbf{Q})=[U+2U^2\Pi_+(\mathbf{Q})]\delta_{l,0}\delta_{l',0} \notag \\
    &\!\!+\! U^2 \!\!\int_{\mathbf{k}}\!\int_{\mathbf{k}'}\!\!f_l(\mathbf{k})f_{l'}^\ast(\mathbf{k}') [\Pi_+(\mathbf{k}\!-\!\mathbf{k}')\!-\!\Pi_-(\mathbf{k}\!+\!\mathbf{k}'\!+\!\mathbf{Q})]\,. \label{eq:Veffll}
\end{align}
Through the inclusion of ph-c and pp corrections, the sign of the effective interaction in the ph-d channel for $l,l'>0$ can be changed 
so that a charge instability with a higher lattice harmonic becomes possible. Without nesting, this could, in principle, lead to a Pomeranchuk instability characterized by $Q=0$. However with nesting, the leading instability is towards an unconventional charge density wave. 

We note that generally neither $\hat{V}_{\mathrm{ph-d}}^{0,ll'}(\mathbf{Q})$ nor $\hat{\Pi}_+^{ll'}(\mathbf{Q})$ are diagonal when expanding in lattice harmonics because of the lower symmetry at finite $Q\neq0$. This means that admixtures of different components within the same irreducible representation of the extended point group are allowed by symmetry \cite{Venderbos2016}. To determine if there is an instability in the ph-d channel, we have to calculate the eigenvalues of the matrix product $\hat{V}_{\mathrm{ph-d}}^{0,ln}(\mathbf{Q})\hat{\Pi}_+^{nl'}(\mathbf{Q})$.  

%--------
\subsection{Beyond KL}
%--------

Going beyond the KL-like analysis to higher order in the bare interaction, the question arises which diagrams to select for resummation to determine the effective interaction in the ph-d channel. 
The diagrams leading in $N_f$ are included in the ph-d channel. 
On the next level, ph-c and pp contributions are of the same order. 
One choice is to only take the summed crossed particle-hole series which is near-singular if the system is at a temperature close to the magnetic instability that arises due to nesting. 
Such an approach is analogous to a spin-fluctuation mechanism for pairing. 
For a constant, bare interaction, it amounts to
\begin{align}
V_{\mathrm{ph-c}}(\mathbf{k}_1-\mathbf{k}_3)=\frac{U}{1-U\,\Pi_+(\mathbf{k}_1-\mathbf{k}_3)}\,.
\end{align}
Taking $V_{\mathrm{ph-c}}$ as effective interaction in the ph-d channel, we obtain
\begin{align}
\hat V_{\mathrm{ph-d}}^{0,ll'}&=\int_{\mathbf{k}}\int_{\mathbf{k}'}f_l(\mathbf{k})f^\ast_{l'}(\mathbf{k}')V_{\mathrm{ph-c}}(\mathbf{k}-\mathbf{k}') \label{eq:Veffspinfluc}\,,
\end{align}
and we note that $V_{\mathrm{ph-d}}^{0,ll'}= V_{\mathrm{ph-d}}^{0,l}\delta_{l,l'}$ is diagonal in this spin-fluctuation approach because there are no particle-particle contributions (see Appendix \ref{sec:ffmixing}).

It is instructive to compare this to the spin-fluctuation approach for pairing, where the effective interaction in the pp channel becomes
\begin{align}
\hat V_{\mathrm{pp}}^{0,ll'}(\mathbf{Q})&=\int_{\mathbf{k}, \mathbf{k}'}f_l(\mathbf{k})f^\ast_{l'}(\mathbf{k}') V_{\mathrm{ph-c}}(\mathbf{k}+\mathbf{k}'-\mathbf{Q})\,.
\end{align}
Typically, a pairing instability occurs for zero total momentum $Q=0$. In this case, there is a direct analogy between the effective interaction in pp and ph-d channels
\begin{align}
\hat V_{\mathrm{ph-d}}^{0,ll'}= \pm  \hat V_{\mathrm{pp}}^{0,ll'}(0), 
\end{align}
where the upper (lower) sign is for $l,l'$ even (odd). 

Thus, we find that, if spin fluctuations induce an attraction in an even (singlet) pairing channel, they also induce an attraction in the corresponding unconventional CDW channel. 
For odd (triplet) channels, it is the other way around: an attraction for pairing  translates to a repulsion for CDW and vice versa. 
We note, however, that in contrast to the instability analysis in the pairing channel, the polarization $\hat\Pi^{l,l'}_+(\mathbf{Q})$ that enters the renormalization of the ph-d vertex is not necessarily diagonal, and one has to diagonalize $\hat V_{\mathrm{ph,d}}^{0,l}\hat\Pi^{l,l'}_+(\mathbf{Q})$ to diagnose the instability.

In the spin-fluctuation approach, we focus on the ladder series in the ph-c channel because it is almost singular. 
However, it is unclear whether 
we can neglect pp diagrams because they are of the same size as the ph-c diagrams in a situation with nesting. 
Another option to explore this issue is an FRG or PRG analysis which takes these contribution into account on equal footing. More generally, in the situation we consider, there is also a strong competition of orders:
SDW, CDW, and pairing channels all possess the same logarithmic singularity. 
Usually, the coupling in the spin channel $U_s=U(k_1,k_2,k_4)$ is attractive on the bare level, giving SDW instabilities a head start. In addition, if fluctuations can turn the coupling in a CDW channel, there is also an attraction in a pairing channel. Assuming $\Pi_{-}(0)\sim\Pi_{+}(Q)$, the condition for a CDW instability to win this competition is
\begin{align}
N_f V_{\mathrm{ph,d}}^{0}>V_{\mathrm{pp}}^{0}, U_s\,,
\end{align} 
which is always possible for large enough $N_f$. Furthermore, an unconventional CDW can benefit from a suppression of the bare interaction in the spin channel. For example, this can happen in systems with several orbitals/sublattices per unit cell  due to sublattice interference effects. ~\cite{PhysRevB.86.121105}. 

The KL description focuses on a minimal subset of the interaction corrections that are crucial for unconventional charge density wave (uCDW) instabilities in order to find a qualitative picture of the working mechanism. We explicitly analyze one-loop spin and pairing corrections to the charge (direct particle-hole) channel to demonstrate how they can induce uCDW states. The inclusion of higher-order corrections improves the description in terms of inter-channel feedback and the perturbative accuracy. To go even further and include all channels on equal footing and resolve their competition depending on the energy scale, a renormalization group approach such as FRG or PRG is appropriate. To resolve sufficient harmonics, we need a momentum resolution that includes the whole Brillouin zone. While patch models, often used in PRG approaches, usually correctly resolve the leading instabilities in a controlled approximation, they cannot account for inversion-odd harmonics or harmonics with an angular momentum higher than the number of patches (see Appendix \ref{app:patchmodel}). Therefore, we choose the truncated unity FRG over a PRG ansatz below. 

In the next section, we show how the KL-like mechanism indeed facilitates instabilities towards unconventional charge order in square and triangular lattices. 
Furthermore, we discuss a spin-fluctuation approach and study a mechanism based on a nearest-neighbor exchange coupling in a phenomenological model with the aforementioned TUFRG. 

%--------
\section{Application to square and triangular lattice}\label{sec:applylattice}
%--------

%--------
\subsection{Lattice models}
%--------

We apply the general formalism developed in Sec.~\ref{sec:KLCDW} to the Hubbard model on the square and triangular lattice with perfect nesting and SU($N_f$) flavour symmetry. To this end, we consider the Hamiltonian $\mathcal{H}=\mathcal{H}_0+ \mathcal{V}$, where 
\begin{align}
    \mathcal{H}_0 &= -t\sum_{\braket{i,j}}\sum_\sigma \left( c_{i,\sigma}^\dagger c_{j\sigma}^{\phantom{\dagger}} + \text{h.c.}\right)-\mu \sum_i \sum_{\sigma}n_{i,\sigma} \label{eq:H0} \\
   \mathcal{V} &=\frac{U}{2}\sum_i\sum_{\sigma,\sigma'} n_{i,\sigma} n_{i,\sigma'}\label{eq:V}
\end{align}
with nearest neighbor hopping $t$, chemical potential $\mu$, and on-site Coulomb repulsion $U>0$. The density operator is given by $n_{i,\sigma}=c_{i,\sigma}^\dagger c_{i\sigma}^{\phantom{\dagger}}$ and the flavour index $\sigma$ runs through $1\leq\sigma \leq N_f$. 
The dispersion relations from diagonalising $\mathcal{H}_0$ on the square and triangular lattice are given by 
\begin{align}
    \epsilon_{\mathbf{k}}^\square &\!=\! -2t \left(\cos(k_x)+\cos(k_y)\right)- \mu\,, \\
    \epsilon_{\mathbf{k}}^\triangle &\!=\! -2t \left(\cos(k_y)\!+\!2\cos(k_y/2)\cos(\sqrt{3}kx/2)\right)\!-\!\mu\,,
\end{align}
where we used nearest-neighbor vectors $\mathbf{a}_1=(1,0)$ and $\mathbf{a}_2=(0,1)$ for the square and $\mathbf{a}_1=(0,1),\mathbf{a}_2=(-\sqrt{3},-1)/2$, and $\mathbf{a}_3=(\sqrt{3},-1)/2$ for the triangular lattice.

We focus on the regimes  around %close to
Van Hove filling for square and triangular lattice, i.e., we %set 
 vary the chemical potential $\mu$ %close to
 around $\mu_{\mathrm{VH}}^\square = 0$ and $\mu_{\mathrm{VH}}^\triangle = 2t$, respectively. 
At %these 
 Van Hove filling , the density of states possesses a logarithmic singularity and the Fermi surface is nested $\epsilon_{\mathbf{k}+\mathbf{Q}}=-\epsilon_\mathbf{k}$. 
At $\mu_{VH}^\square$, nesting is perfect on the square lattice with $\mathbf{Q}^\square =(\pm \pi,\pm \pi)$. 
On the triangular lattice  at $\mu_{VH}^\triangle$, %it is approximate around the Fermi surface with 
 there are three nesting vectors $\mathbf{Q}^\triangle \in \{M_1,M_2,M_3\}$ and
$M_1=(\pm 2\pi/\sqrt{3},0), M_2=\pm (\pi/\sqrt{3},\pi),  M_3=\pm (\pi/\sqrt{3},-\pi)$  which connect opposite sides of the hexagonal Fermi surface. Tuning $\mu$ away from Van Hove energy, nesting of the Fermi surface is destroyed. 
Under these conditions, there is a double logarithmic singularity in both the particle-particle and the particle-hole polarization $\Pi_-(0)\sim\Pi_+(\mathbf{Q})\sim\ln^2(W/\text{max}(T,\delta\mu))$ with bandwidth $W$, temperature $T$, and distance to Van Hove filling $\delta\mu=|\mu-\mu_{\mathrm{VH}}|$.
Thus, the prerequisites outlined in Sec.~\ref{sec:KLCDW} for an instability towards an unconventional CDW are satisfied for sufficiently large $N_f$. 

%--------
\subsection{Kohn-Luttinger-like mechanisms applied}
%--------

We first consider CDW instabilities with wave vector $\mathbf{Q}=\mathbf{Q}^\square$ or $\mathbf{Q}=\mathbf{Q}^\triangle$ on the square and triangular lattice, respectively. 
In each case, we numerically calculate the form-factor expansion of the effective interaction that we use as input for the RPA summation of the ph-d channel according to Eq.~\eqref{eq:Veffll}. 
This means we include one-loop pp and ph-c corrections, as well as the constant ph-d one-loop corrections which are a factor of $N_f$ smaller. 
Similarly, we numerically calculate and expand the polarization operator $\hat \Pi^{l,l'}_+(\mathbf{Q})$. 
We use the numerical error estimates to round the results. 

For our calculation, we use the lattice harmonics for the square and triangular lattices as a form-factor basis.  
We truncate the expansion after nearest neighbors and check that including next-nearest neighbors does not qualitatively change the leading CDW instability (see Appendix \ref{app:ff} for the explicit expressions.) 
For the square lattice this amounts to the inclusion of $s$-, $s'$-, $d_{x^2-y^2}$-, $p_x$, and $p_y$-wave nearest-neighbor form factors. 
For the triangular lattice it corresponds to the inclusion of $s$-, $s'$-, $f_{y(y^2-3x^2)}$-, $d_{x^2-y^2}$-, $d_{xy}$-, $p_x$, and $p_y$-wave form factors. 
Interestingly, we find that in the considered cases pp and ph-c fluctuations work against each other for even harmonics, while they work together for odd harmonics. 
To show this, we rewrite for $l,l'>0$
\begin{align}
   &\hat V_{\text{ph-d}}^{0,l,l'}(\mathbf{Q})= U^2 \int_{\mathbf{k}}\int_{\mathbf{k}'}f_l(\mathbf{k})f_{l'}^\ast(\mathbf{k}') \notag \\
    &\quad\quad\quad\times[\Pi_+(\mathbf{k}-\mathbf{k}')-(-1)^{l'}\Pi_-(\mathbf{k}-\mathbf{k}'+\mathbf{Q})]\, . \label{eq:Veffll2}
\end{align}
using that $f_l(-\mathbf{k})=(-1)^l f_l(\mathbf{k})$. 
On the square lattice, $\Pi_-(\mathbf{k}-\mathbf{k}'+\mathbf{Q}^\square)=\Pi_+(\mathbf{k}-\mathbf{k}')$ due to perfect nesting. It follows that $\hat V^{0,l,l'}_{\mathrm{ph-d}}(\mathbf{Q}^\square)=0$ for even $l'$, and $\hat V^{0,l,l'}_{\mathrm{ph-d}}(\mathbf{Q}^\square)=2U^2\int_\mathbf{k}\int_{\mathbf{k}'}f_l(\mathbf{k})f_{l'}^\ast(\mathbf{k}')\Pi_+(\mathbf{k}-\mathbf{k}')$ for odd $l'$. On the triangular lattice, nesting is not perfect, however the leading logarithmic behavior still obeys $\Pi_-(\mathbf{k}-\mathbf{k}'+\mathbf{Q}^\triangle)\approx\Pi_+(\mathbf{k}-\mathbf{k}')$, i.e., to a good approximation these contributions also cancel for even~$l'$ and add up for odd~$l'$. 

\begin{figure}
    \centering   \includegraphics[width=0.9\columnwidth]{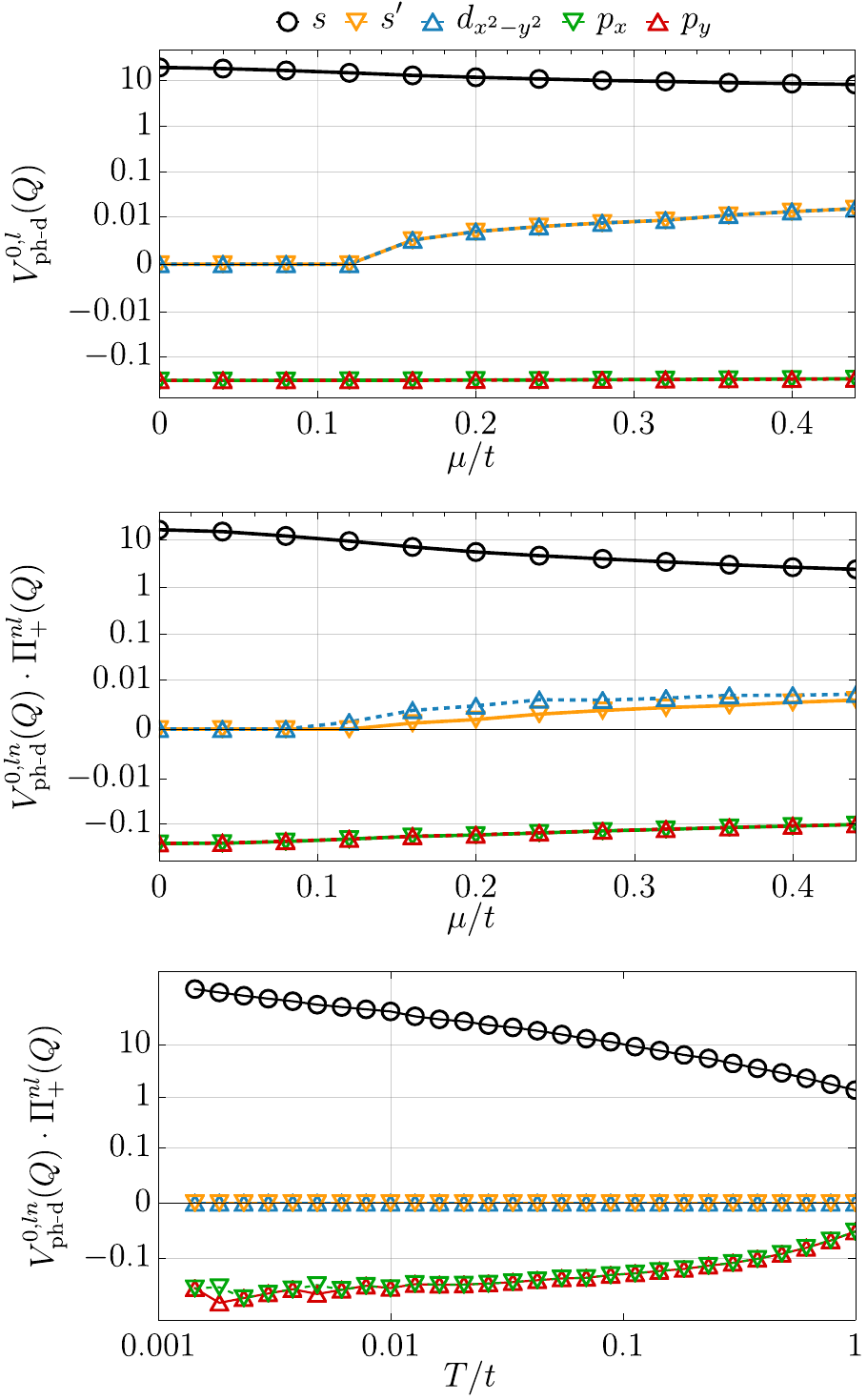}
    \caption{\textbf{Square lattice.} (Top) Effective interaction in the ph-d channel $V_{\text{ph-d}}^{0,l}(\mathbf{Q}^\square)$ including Kohn-Luttinger-like corrections, expanded in nearest-neighbor lattice harmonics $l\in\{s,s',p_x,p_y,d_{x^2-y^2}\}$ as a function of the chemical potential $\mu$ at temperature $T=0.05/t$ and bare interaction $U=3t$.
    (Middle) Product of effective interaction and polarization $V_{\text{ph-d}}^{0,l}(\mathbf{Q}^\square)\Pi_+^l(\mathbf{Q}^\square)$ as it enters the direct particle-hole channel, cf. Eq.~\eqref{eq:phd_ladder_frac}, i.e., the most negative one determines the CDW instability.
    (Bottom) Eigenvalues of $V_{\text{ph-d}}^{0,l}(\mathbf{Q}^\square)\Pi_+^l(\mathbf{Q}^\square)$ as a function of temperature $T/t$ at $\mu=\mu_{VH}$.
    }
     \label{fig:Veffsquare}
\end{figure}

%--------
\subsubsection{Square lattice}
%--------

In the square-lattice case, both the effective interaction and the polarization operator decouple between the different irreducible representations of the lattice point group for $\mathbf{Q}=\mathbf{Q}^\square$, see Appendix \ref{sec:ffmixing}. 
This means that they are diagonal in our truncated form-factor basis $\hat V_{\text{ph-d}}^{0,l,l'}(\mathbf{Q}^\square)=V_{\text{ph-d}}^{0,l}(\mathbf{Q}^\square)\delta_{l,l'}$ and $\hat \Pi_+^{l,l'}(\mathbf{Q}^\square)=\Pi_+^{l}(\mathbf{Q}^\square)\delta_{l,l'}$, except for $A_1$ representatives, i.e., these relations hold for $l,l'\notin A_1$. 
Furthermore, all components of the polarization operator are positive so that it suffices to analyze the sign of $V_{\text{ph-d}}^{0,l}(\mathbf{Q}^\square)$ to determine if there is an instability towards an unconventional CDW.

We show the result in Fig.~\ref{fig:Veffsquare}. We find that the $p_x$ and $p_y$ components become attractive so that an instability in the $p$-wave CDW channel arises. This means that the singular part of the effective interaction is of the form  
\begin{align}
V=\sum_{k,\sigma}\sum_{\alpha=x,y}p_\alpha(\mathbf{k})p_\alpha(\mathbf{k}')c_{\mathbf{k} \sigma}^\dagger
    c_{\mathbf{k}'+\mathbf{Q} \sigma'}^\dagger
    c_{\mathbf{k}' \sigma'}
    c_{\mathbf{k}+\mathbf{Q} \sigma}\,,
\end{align}
suggesting a mean-field decoupling with order parameter
\begin{align}
    \Delta_\alpha(\mathbf{Q})=\sum_{k,\sigma}p_\alpha(\mathbf{k}) c_{\mathbf{k}+\mathbf{Q} \sigma}^\dagger
    c_{\mathbf{k} \sigma'}\,.
\end{align}
The resulting state of broken symmetry corresponds to a bond density wave with alternating strong and weak bonds in $x$- and $y$ directions, breaking inversion symmetry. 
The exact pattern of the bond density wave depends on the linear combination of $p_x$ and $p_y$ in the ground state. 
It can be alternating in just one of the directions or, if both contributions are of the same order, it yields a zig-zag bond pattern. 

We also find that at Van Hove filling, pp and ph-c contributions exactly cancel for $d$-wave and extended $s$-wave components $V_{\text{ph-d}}^{0,d}(\mathbf{Q}^\square)=V_{\text{ph-d}}^{0,s'}(\mathbf{Q}^\square)=0$ as expected from Eq.~\eqref{eq:Veffll2} due to perfect nesting. 
We confirm this observation for the full model also by an effective low-energy patch model that only keeps states around Van Hove points, see Appendix \ref{app:patchmodel}. 
Away from Van Hove filling $\Pi_-(0)>\Pi_+(\mathbf{Q}^\square)$ and the coupling in $s'$- and $d$-wave CDW channels becomes repulsive, see Fig.~\ref{fig:Veffsquare}. 

%--------
\subsubsection{Triangular lattice}
%--------

\begin{figure}
    \centering
\includegraphics[width=0.9\columnwidth]{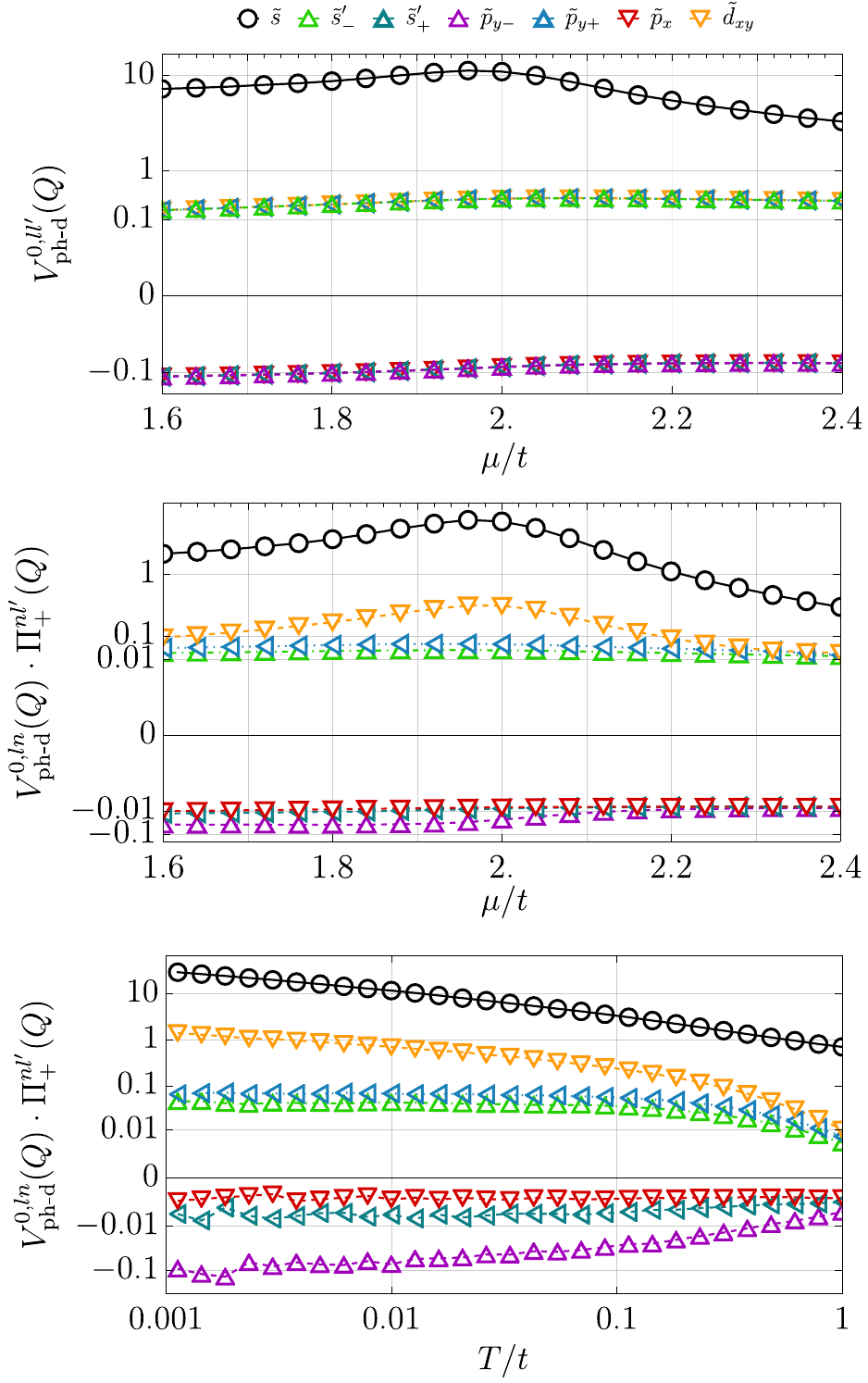}
    \caption{\textbf{Triangular lattice.} (Top) Eigenvalues of $\hat V_{\text{ph-d}}^{0}(M_1)$ including Kohn-Luttinger-like corrections expanded in nearest-neighbor lattice harmonics $l\in\{s,s',p_x,p_y,d_{x,y},d_{x^2-y^2},f_{y(y^2-3x^2)}\}$ as a function of the chemical potential $\mu$ at temperature $T=0.05/t$ and bare interaction $U=3t$. The eigenvectors are linear combinations of the lattice harmonics and labeled according to the lower symmetry at $\mathbf{Q}^\triangle$ 
    (see Appendix \ref{app:ff} for explicit expressions). (Middle) Eigenvalues of $\hat V_{\text{ph-d}}^{0}(M_1)\hat \Pi_+(M_1)$ which enter the direct particle-hole channel, cf. Eq.~\eqref{eq:phd_ladder_frac}, i.e., the most negative one determines the CDW instability.
    (Bottom) Eigenvalues of $\hat V_{\text{ph-d}}^{0}(M_1)\hat \Pi_+(M_1)$ as a function of temperature $T/t$ at $\mu=\mu_{VH}$.}
     \label{fig:Vefftri}
\end{figure}

For the triangular lattice, the analysis is slightly more involved. 
In this case, the KL-like interaction $\hat V_{\text{ph-d}}^{0,l,l'}(\mathbf{Q}^\triangle)$ and the polarization $\hat \Pi_+^{l,l'}(\mathbf{Q}^\triangle)$ are not diagonal because of the lower two-fold symmetry at $\mathbf{Q}^\triangle$. 
Mixing between different nearest-neighbor form factors is now allowed because they fall into the same irreducible representations of the little group at $\mathbf{Q}^\triangle$.
For example, if $Q^\triangle=M_1$ mixing can occur between $s$, $s'$- and $d_{x^2-y^2}$- and between $f_{y(y^2-3x^2)}$- and $p_y$-basis functions. 
To analyze if an instability occurs in the ph-d channel, we determine the eigenvalues and eigenfunctions of $\hat V_{\text{ph-d}}^{0} \cdot \hat \Pi_+$.
We find three attractive channels with negative eigenvalues as shown in Fig.~\ref{fig:Vefftri}. 
They correspond to $p$- and extended $s$-wave form factors for the little group, which we obtain through linear combinations of the nearest-neighbor harmonics of the triangular lattice (see Appendix \ref{app:ff}). 
The strongest attraction occurs in the $\tilde p_{y-}$ channel
\begin{align}
    \tilde p_{y-}&=\frac{1}{\sqrt{3}}(p_y-\sqrt{2}f)\notag\\
    &=\sin (\mathbf{a}_2\cdot \mathbf{k})-\sin (\mathbf{a}_3\cdot \mathbf{k})
\end{align}
An analogous analysis applies to $Q^\triangle\in\{M_2,M_3\}$, which can be obtained from the expressions for $M_1$ by rotation of $\pm\pi/3$. 
This suggests an instability towards a bond density wave described by three possible order parameters $\Delta(M_1),\Delta(M_2),\Delta(M_3)$ with 
\begin{align}
    \Delta(M_1)=\sum_{k,\sigma}[\sin (\mathbf{a}_2\cdot \mathbf{k})-\sin (\mathbf{a}_3\cdot \mathbf{k})] c_{\mathbf{k}+\mathbf{Q} \sigma}^\dagger
    c_{\mathbf{k} \sigma'}\,
\end{align}
and analogously for $\Delta(M_2),\Delta(M_3)$. The resulting bond density wave for $\Delta(M_1)\neq 0$ has a zig-zag pattern that breaks inversion with alternating bonds along $\mathbf{a}_2$ and $\mathbf{a}_3$. The pattern in the ground state depends on the linear combination of all three $\Delta(M_i)$. 

%--------
\subsection{Spin-fluctuation approach}
%--------

Away from large $N_f$, diagrams beyond the KL-like approach need to be included. 
We discussed two options to do so in Sec.~\ref{sec:KLCDW}, a spin-fluctuation and a functional renormalization group (FRG) approach. 
In the spin-fluctuation approach, the effective interaction in pp and ph-d channels are related via $\hat V_{\mathrm{ph,d}}^{0,l}= \pm  \hat V_{\mathrm{pp}}^{0,l}(0)$ (both are diagonal in $l$). 
Thus, we can utilize earlier studies of the spin-fluctuation mechanism for pairing in the square and triangular lattice \cite{PhysRevB.34.8190,PhysRevB.51.3797,PhysRevB.81.224505} to extract the effective interaction in the ph-d channel. 
We conclude that an instability towards a $d_{x^2-y^2}$ CDW occurs on the square lattice as $d_{x^2-y^2}$ superconductivity has been established as the leading pairing instability there \cite{RevModPhys.84.1383}.

The resulting state of broken symmetry describes a loop-current order with a staggered-flux pattern on neighboring plaquettes. 
This is also consistent with PRG and FRG studies, which find the staggered-flux state for $N_f\geq 4$ \cite{Lin2019} and $N_f\geq 6$ \cite{Honerkamp2004p1}, respectively. 
On the triangular lattice, we can infer an attraction in the
$d$-wave CDWs channel based on earlier analyses \cite{PhysRevB.81.224505}. 
We note that the $f$-wave pairing state which is competitive on the triangular lattice, becomes repulsive in the CDW channel because of the additional minus sign for odd harmonics in the relation between the couplings in pairing and ph-d channel. 
This suggests that an instability towards a $d$-wave CDW can occur. However, for a proper treatment, the eigenvalues of $\hat V_{\mathrm{ph-d}}^0\hat \Pi_+$ need to be determined (see next Sec.~\ref{sec:tju}).

The $d$-wave CDW state can leads to various loop current and flux patterns depending on the combination of the three degenerate CDW with wave vectors $M_1,M_2,M_3$. 
For a single $M$ point, an alternating current pattern along two directions of the triangular lattice is formed, and for triple-M order a quantum anomalous Hall state that quadruples the unit cell appears~\cite{Venderbos2016}. 

%--------
\subsection{Instabilities without nesting}
%--------

If we depart from a weak-coupling analysis, unconventional CDWs can arise as Stoner instabilities for large $N_f$ and appropriate bare momentum-dependent interactions even without nesting. 
As an example, we consider a phenomenological $t-U-J$ model on the square and triangular lattice  $\mathcal H\rightarrow\mathcal H_0+\mathcal V+ \mathcal H_J$, where 
\begin{align}
    \mathcal H_J=J \sum_{\langle i,j \rangle} \sum_{\alpha\ldots\delta}\sum_{n}(c_{i\alpha}^\dagger T^n_{\alpha\beta} c_{i\beta}) (c_{j\gamma}^\dagger T^n_{\gamma\delta}c_{j\delta})\,,
\end{align}
with the SU($N_f$) generators $T^n$ we used previously in Eq.~\eqref{eq:chargespinvertex} and $\mathcal H_0$, $\mathcal V$ are given in Eqs.~\eqref{eq:H0} and~\eqref{eq:V}.

After a Fourier transform, we expand the exchange interaction into form factors in the pairing $U_{\mathrm{pp}}$ ($k_1=-k_2=k,k_3=k'$), ph-c $U_{\mathrm{ph-c}}$ ($k_1=k,k_2=k'+Q,k_3=k+Q$) and ph-d $U_{\mathrm{ph-d}}$ ($k_1=k,k_2=k'+Q,k_3=k'$) channels.
Explicitly, on the square lattice, we then obtain 
\begin{align}
    &U_{\mathrm{pp}}(\mathbf{k},\mathbf{k}',0)=U \notag \\
    &-\frac{J}{2}\left(1+\frac{1}{N_f}\right) [s'(\mathbf{k})s'(\mathbf{k'})+d_{x^2-y^2}(\mathbf{k})d_{x^2-y^2}(\mathbf{k'})]\notag \\
    &-\frac{J}{2}\left(1-\frac{1}{N_f}\right) [p_x(\mathbf{k})p_x(\mathbf{k'})+p_y(\mathbf{k})p_y(\mathbf{k'})]\,,\\[5pt]
    &U_{\mathrm{ph-c}}(\mathbf{k},\mathbf{k}',\mathbf{Q})=U-\frac{J}{4} s'(\mathbf{Q}) \notag \\
    &-\frac{J}{2N_f}[s'(\mathbf{k})s'(\mathbf{k'}) +p_x(\mathbf{k})p_x(\mathbf{k'})
    +p_y(\mathbf{k})p_y(\mathbf{k'}) \notag\\
    &\qquad\quad+ d_{x^2-y^2}(\mathbf{k})d_{x^2-y^2}(\mathbf{k'})]\,,\\[5pt]
    &U_{\mathrm{ph-d}}(\mathbf{k},\mathbf{k}',\mathbf{Q})=U-\frac{J}{4N_f}s'(\mathbf{Q}) \notag \\
    &-\frac{J}{2}[s'(\mathbf{k})s'(\mathbf{k'}) +p_x(\mathbf{k})p_x(\mathbf{k'})
    +p_y(\mathbf{k})p_y(\mathbf{k'})\notag\\
    &\qquad+ d_{x^2-y^2}(\mathbf{k})d_{x^2-y^2}(\mathbf{k'})]\,,\label{eq:Jsquare}
\end{align}
and on the triangular lattice, we find 
\begin{align}
    &U_{\mathrm{pp}}(\mathbf{k},\mathbf{k}',0)=U \notag \\
    &-\frac{J}{2}\left(1+\frac{1}{N_f}\right) [s'(\mathbf{k})s'(\mathbf{k'})
    + d_{xy}(\mathbf{k})d_{xy}(\mathbf{k'}) \notag \\
    &\hspace{2.5cm} + d_{x^2-y^2}(\mathbf{k})d_{x^2-y^2}(\mathbf{k'})]\notag \\
    &-\frac{J}{2}\left(1-\frac{1}{N_f}\right) [p_x(\mathbf{k})p_x(\mathbf{k'})+p_y(\mathbf{k})p_y(\mathbf{k'}) \notag \\
    & \hspace{2.5cm}+ f_{y(y^2-3x^2)}(\mathbf{k})f_{y(y^2-3x^2)}(\mathbf{k'})]\,,\\[5pt]
    &U_{\mathrm{ph-c}}(\mathbf{k},\mathbf{k}',\mathbf{Q})=U-\sqrt{\frac{3}{2}}J s'(\mathbf{Q}) \notag \\
    &-\frac{J}{2N_f}[s'(\mathbf{k})s'(\mathbf{k'})+p_x(\mathbf{k})p_x(\mathbf{k'})
    +p_y(\mathbf{k})p_y(\mathbf{k'})\notag\\
    &\qquad\quad + d_{x^2-y^2}(\mathbf{k})d_{x^2-y^2}(\mathbf{k'})+ d_{xy}(\mathbf{k})d_{xy}(\mathbf{k'}) \notag \\
    &\qquad\quad+ f_{y(y^2-3x^2)}(\mathbf{k})f_{y(y^2-3x^2)}(\mathbf{k'})]\,,\\[5pt]
    &U_{\mathrm{ph-d}}(\mathbf{k},\mathbf{k}',\mathbf{Q})=U-\sqrt{\frac{3}{2}}\frac{J}{N_f}s'(\mathbf{Q}) \notag \\
    &-\frac{J}{2}[s'(\mathbf{k})s'(\mathbf{k'}) +p_x(\mathbf{k})p_x(\mathbf{k'})
    +p_y(\mathbf{k})p_y(\mathbf{k'})\notag\\
    &\qquad+ d_{x^2-y^2}(\mathbf{k})d_{x^2-y^2}(\mathbf{k'})+ d_{xy}(\mathbf{k})d_{xy}(\mathbf{k'}) \notag \\
    &\qquad+ f_{y(y^2-3x^2)}(\mathbf{k})f_{y(y^2-3x^2)}(\mathbf{k'})]\,. \label{eq:Jtriangle}
\end{align}
At these expressions, we can read off the components of the bare interaction in the form-factor expansion, which are all diagonal $\hat U^{l,l'}_{X}=\hat U^{l}_{X}\delta_{l,l'}$, $X\in$ \{pp,ph-c,ph-d\}. We see that the bare interaction is attractive in the ($s$-wave) spin channel, the pairing channel, and the charge channel for all nearest-neighbor harmonics. 
We note in passing that the instability in the spin channel is determined by the maximum value of $[U-c_iJs'(\mathbf{Q})]\Pi_+(\mathbf{Q})$ ($c_\square=1/4$, $c_\triangle=\sqrt{3/2}$), which leads to the expected SDWs with $\mathbf{Q}^\square$ or $\mathbf{Q}^\triangle$. Similarly, the pairing instability with the largest critical temperature is a $d$-wave, because it has a larger interaction and a larger $\Pi_-^l(0)$ than other harmonics. It generally also depends on the filling, which of the spin, pairing, or charge instabilities wins. However,   
for a large enough value of $N_f$, we can ensure that a charge instability wins the competition. 
Its type is determined by the singularities of
\begin{align}
     \hat{V}^{l,l'}_{\text{ph-d}}(\mathbf{Q}) = -2\left[\frac{1}{\mathbb{1} + N_f \hat{U}_{\mathrm{ph-d}} \hat{\Pi}_+(\mathbf{Q})}\right]_{ll'}\hat U_{\mathrm{ph-d}}^{l'}\, \label{eq:VphdfortjU}
\end{align}

\label{sec:tju}
\begin{figure}
        \centering
    \includegraphics[width=\columnwidth]{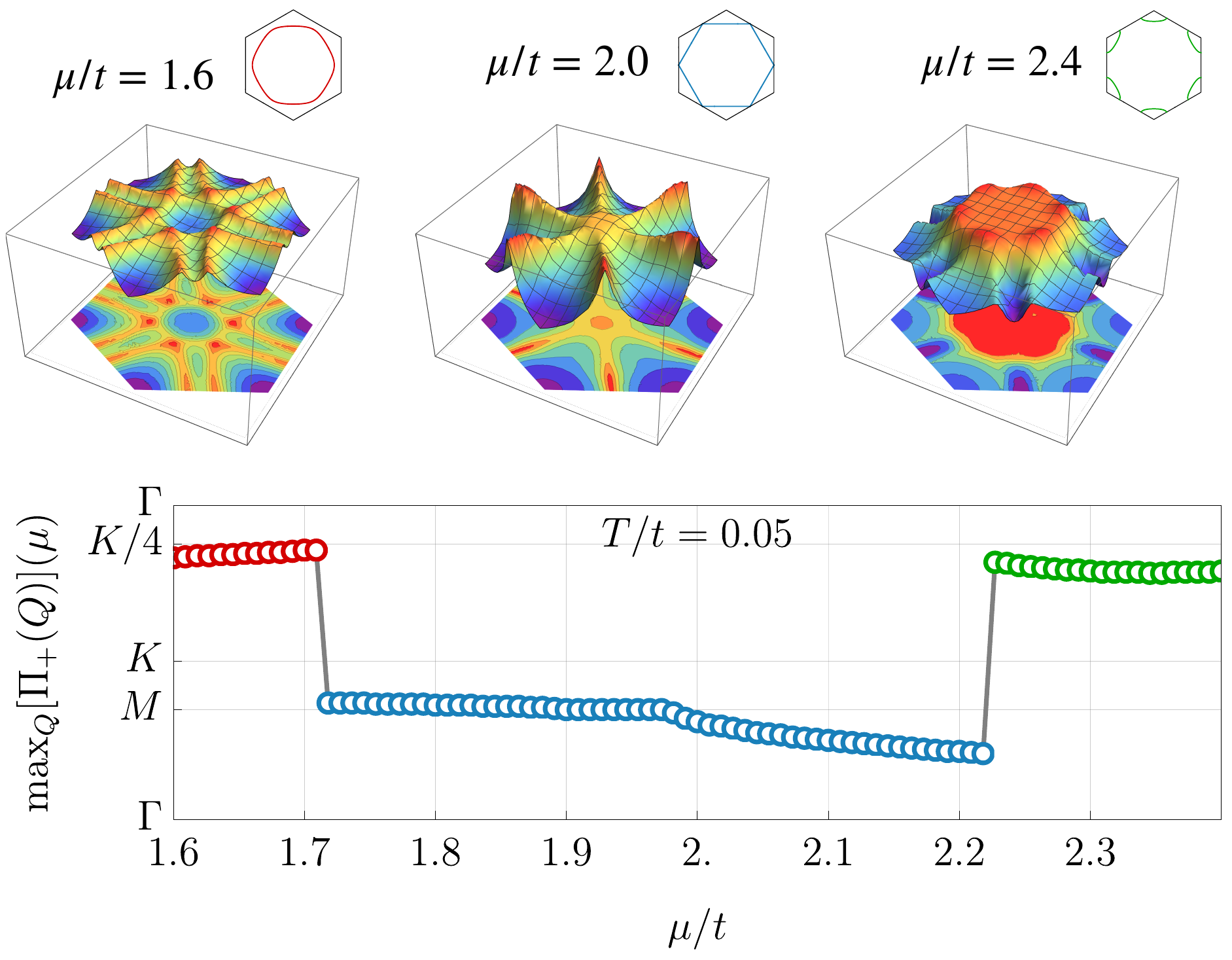}
    \caption{\textbf{Electronic polarization at various fillings} for the triangular lattice. (Top) Bare particle-hole polarization $\Pi_+(\mathbf{k})$ at $\mu=1.6t$ (left), $\mu_{\mathrm{VH}}$ (middle), and $\mu=2.4t$ (right). The position of the largest peak shifts from $M$ to $K/4$. 
    (Bottom left) The position of $Q$ which maximizes the polarization $\Pi_+$. 
    }
    \label{fig:K4}
\end{figure}

\begin{figure}
        \centering       \includegraphics[width=\columnwidth]{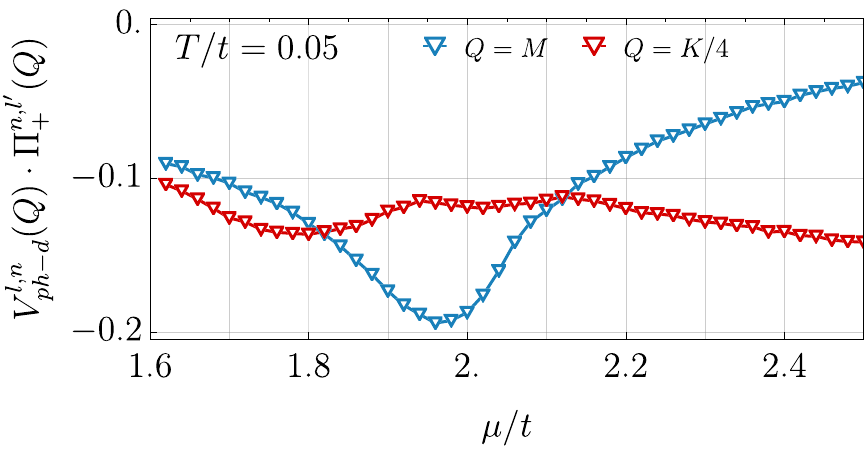}
    \caption{\textbf{Most negative eigenvalue} of $\hat V_{\mathrm{ph-d}}^{ln}(Q)\hat\Pi_+^{nl'}(Q)$ for $Q=M$ (blue) and $Q=K/4$ (red), which determines the uCDW instability, calculated for temperature $T=0.05t$ and interaction $J=t$. The eigenfunctions have a predominant $d$-wave ($f-$wave) form for $\mu< \mu_{c2}$ ($\mu\geq \mu_{c2}$) as described in the Sec.~\ref{sec:tju}. The crossing implies a phase transition where the wave vector changes from~$M$ to~$K/4$.}
    \label{fig:K4CDW}
\end{figure}

On the square lattice, we find that the leading instability at Van Hove filling occurs for $\mathbf{Q}=\mathbf{Q}^\square$, where $\Pi^{l,l'}_+(\mathbf{Q})$ is diagonal and largest for $l=d_{x^2-y^2}$. This is again the staggered flux state and reproduces earlier analyses for large $N_f$ \cite{PhysRevB.37.3774,PhysRevB.39.11538}.

Similarly, the leading instability on the triangular lattice at Van Hove filling is a $d$-wave CDW at $\mathbf{Q}=\mathbf{Q}^\triangle$, where $\mathbf{Q}^\triangle$ is one of the $M$ points. 
For a given $M$ point, the unconventional CDW has a unique $d$-wave form factor. 
To determine it, we diagonalize $\hat U_{\mathrm{ph}} \hat{\Pi}_+(\mathbf{Q})$, which yields a pure $d_{xy}$ at $M_1$, and the appropriately rotated superposition $\cos(\pi/3)d_{x^2-y^2}\pm \sin(\pi/3) d_{xy}$ at $M_{2,3}$. 
This is the alternating or loop current state for single-/triple-M order we mentioned in the previous section. 

Changing the chemical potential slightly away from Van Hove doping, the wave vector where the polarization $\Pi_+(\mathbf{Q})$ peaks moves away from $\mathbf{Q}^\square$ or $\mathbf{Q}^\triangle$ and becomes incommensurate. 
Interestingly, in the triangular lattice, we find that not too far away from Van Hove filling when $\mu<\mu_{c1}\approx 1.8t$ and $\mu>\mu_{c2}\approx 2.1$, the global maximum of $\Pi_+(\mathbf{Q})$ abruptly changes from $\mathbf{Q}\approx\mathbf{Q}^\triangle$ to $\mathbf{Q}\approx K/4$ and symmetry equivalent points, see Fig.~\ref{fig:K4}. 
An analogous change does not occur on the square lattice.  The wave vector of $K/4$ approximately connects the locus of Fermi surface pockets for $\mu>\mu_{c2}$. For $\mu<\mu_{c1}$, $K/4$ corresponds to the wave vector that connects points on the Fermi surface closest to the Van Hove points back-folded into the first Brillouin zone.
 
The change of the peak position of the polarization $\Pi_+$ implies that there is a phase transition as a function of chemical potential $\mu$ where the type of the uCDW changes. We determine the instabilities of $\hat V_{\mathrm{ph}}^{l,l'}(K/4)$ for this case. We find that different CDWs with wave-vector K/4 overtake the $d$-wave CDW with wave-vector M for $\mu<\mu_{c1}$ and $\mu>\mu_{c2}$ as we show in Fig.~\ref{fig:K4CDW}. 
It depends on the chemical potential, in which angular-momentum channel the leading instability occurs. The only symmetry that remains at $K/4$ is a reflection at the $\Gamma$-$K$ line, so we can distinguish form factors being even or odd under this reflection. We find an odd form factor for $\mu<\mu_{c1}$ and an even form factor for $\mu>\mu_{c2}$.
\begin{figure}
    \centering
    \includegraphics[width=\columnwidth]{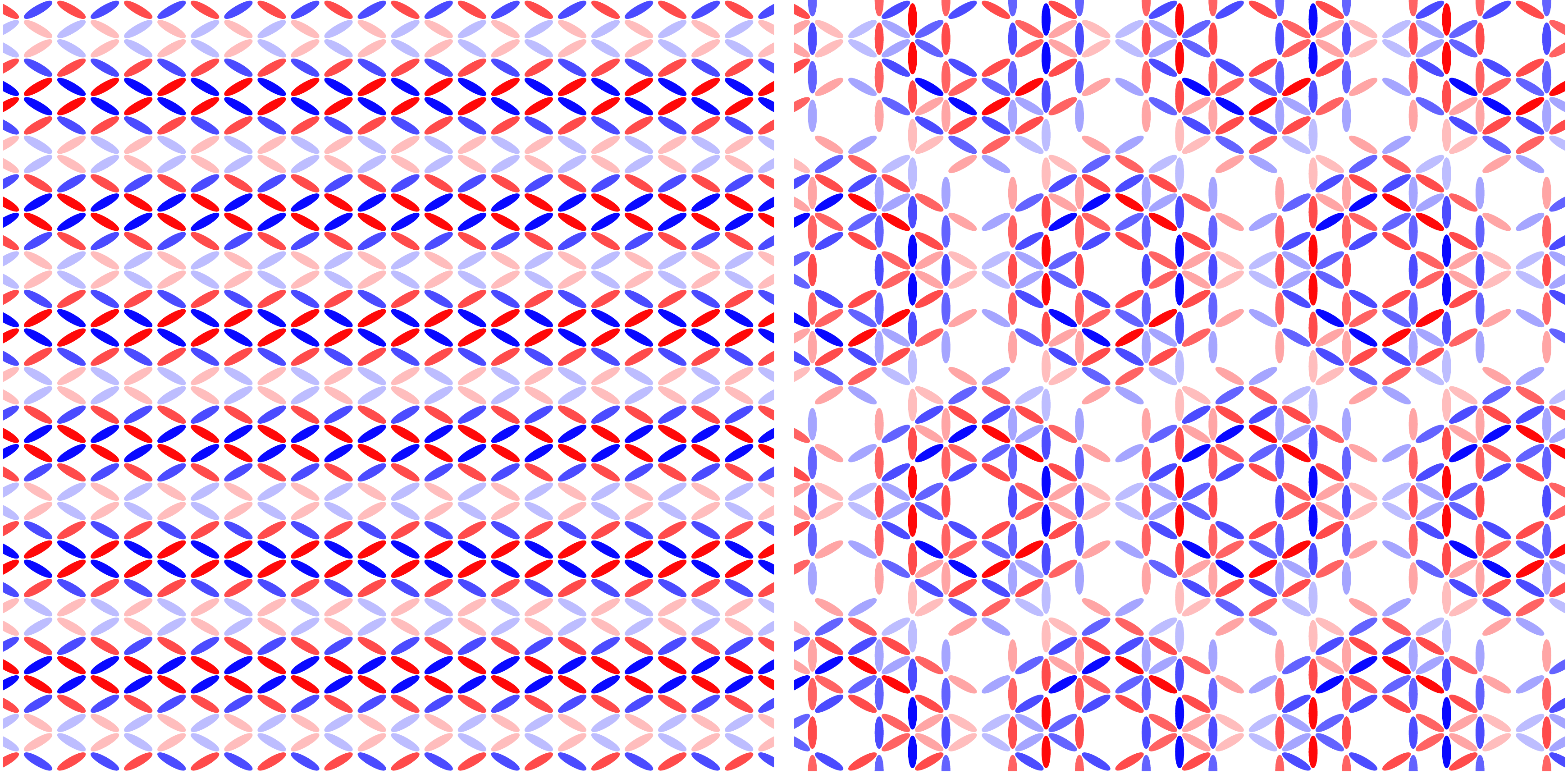}
    \caption{\textbf{The real space pattern} of the reflection-odd ``$p_x+d_{xy}$'' CDW instability for single-Q (left,  $Q=K/4\  \hat e_y$) and triple-Q (right, $Q_i=K_i/4$) configurations. The orders induce imaginary hoppings along bonds, i.e., current patterns, with directions depicted by arrows and amplitudes by the gray scale. }
    \label{fig:realspace_dwave}
\end{figure}
\begin{figure}
    \centering
    \includegraphics[width=\columnwidth]{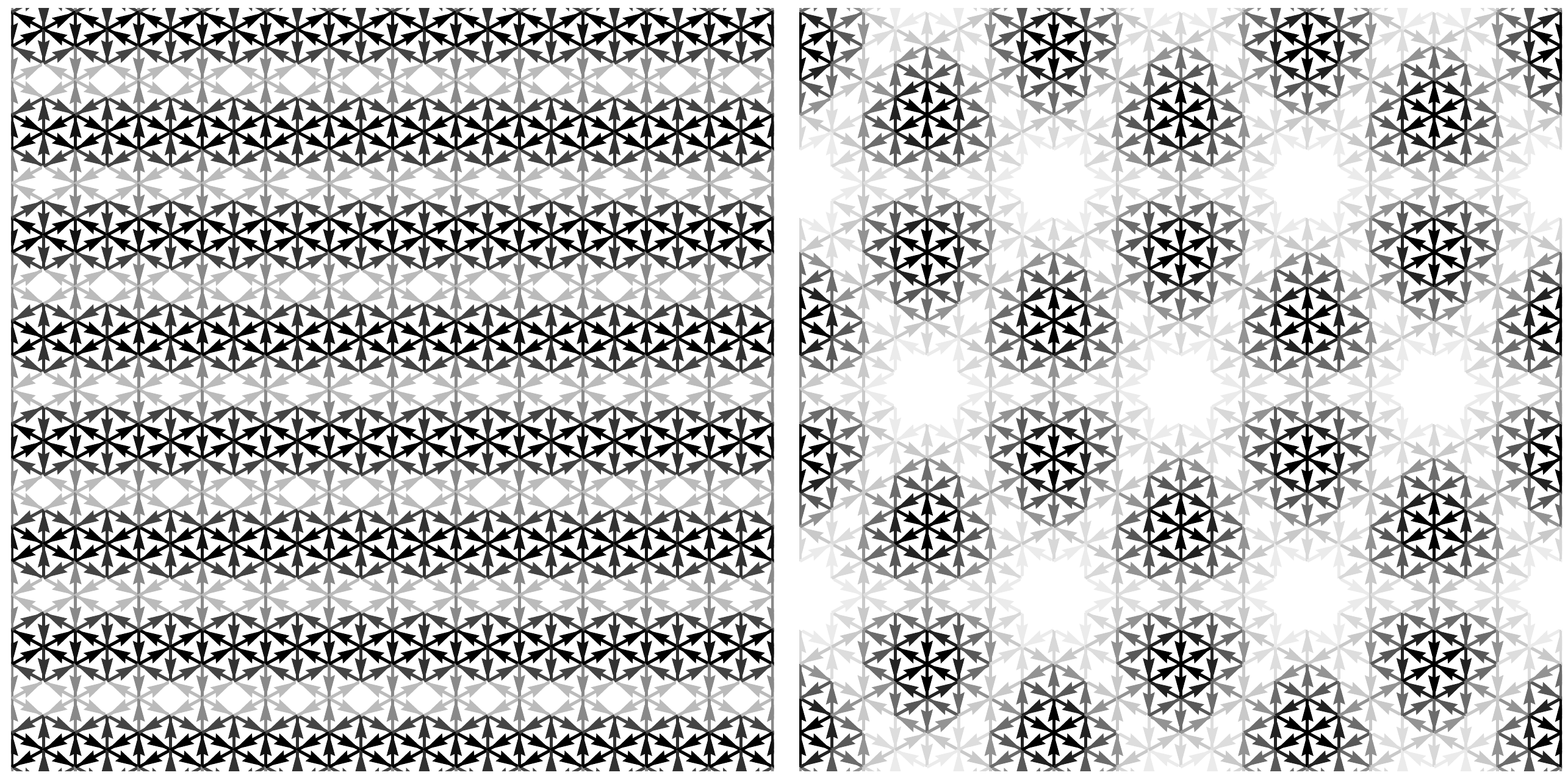}
    \caption{\textbf{The real space pattern} of the reflection-even ``$f_y+d_{x^2-y^2}+ p_y$'' CDW instability for single-Q (left,  $Q=K/4\ \hat e_y$) and triple-Q (right, $Q_i=K_i/4$) configurations. The orders induce imaginary hoppings along bonds, i.e., current patterns, with directions depicted by arrows and amplitudes by the gray scale. }
    \label{fig:realspace_fwave}
\end{figure} 
For $Q=K_1/4=(0,\pi/3)$ and $\mu<\mu_{c1}$,  the odd form factor has a predominant $d_{xy}$-wave form with admixtures from $p_x$, specifically we obtain $\cos (\pi/12) d_{xy}+ \sin(\pi/12)p_x$ within our numerical resolution. This corresponds to a bond order parameter in real space, see Fig.~\ref{fig:realspace_dwave}. Note that the orthogonal linear combination leads to a loop current order, which is, however, subleading.
Similar to the $M$-point instability, the form factors at the six different K/4 points are related by $\pi/3$ rotation. 
For $Q=K_1/4$ and $\mu>\mu_{c2}$, the even form factor we obtain is a superposition of $f$-, $d_{x^2-y^2}$, and (very small) $p_y$-wave ($-0.93 f_y-0.36 d_{x^2-y^2}+0.03 p_y$). %In both cases, the superposition cannot be an arbitrary linear combination and 
The specific linear combination leads to a loop current order, see Fig.~\ref{fig:realspace_fwave}. We checked explicitly that charge conservation on a given site $R$ is satisfied $d\langle n_R\rangle/dt=0$ \cite{Palle2024}, i.e., Kirchhoff's circuit law is fulfilled. 
Furthermore, there can be different linear combinations of degenerate $K_i/4$ density waves and the one with the lowest Landau free energy will form the ground state. 
Examples for the corresponding patterns can be seen in Figs. \ref{fig:realspace_dwave} and \ref{fig:realspace_fwave}.

To show that an uCDW with wave vector K/4 can win against pairing for realistic $N_f$ and upon inclusion of all channels, we perform an FRG calculation for the $t-U-J$ model on the triangular lattice with $N_f=4$. The generalization of the usual SU($2$) FRG equations \cite{Lichtenstein2018,wang2014competing,salmhofer2001fermionic} comes with a single change as only the factor in front of the particle-hole direct term including the fermionic bubble changes from $2$ to the respective $N_f$ \cite{Honerkamp2004p1}.
Within the FRG truncation we employ here (no self-energy feedback, static vertices and only two-particle interactions), the effective vertex is obtained as the solution of the differential equation
\begin{align}\label{eq:FRG}
    \partial_T V = \tau^T_{\mathrm{pp}} + \tau^T_{\mathrm{ph-c}} + \tau^T_{\mathrm{ph-d}}\,,
\end{align}
where temperature is used as the flow parameter and $\tau^T_X$ are given by the expressions in Eqs.~\eqref{eq:tpp}-\eqref{eq:tphd} with the replacement $\mathcal G(k)\rightarrow\mathcal G^T(k)=\sqrt{T}/(i\omega-\xi_\mathbf{k})$ \cite{PhysRevB.64.184516}. The differential form incorporates an all-channel-to-all-channel feedback, i.e., this analysis goes beyond the single channel resummation of Eq.~\eqref{eq:VphdfortjU}. The momentum dependence is resolved via a so-called truncated-unity scheme, which also relies on a form-factor expansion of the effective vertex in the pp, ph-c, and ph-d channels \cite{Lichtenstein2017}. The details of our implementation are described in Ref.~\cite{Gneist2022p2}. 

\begin{figure} 
    \centering
 \includegraphics[width=\linewidth]{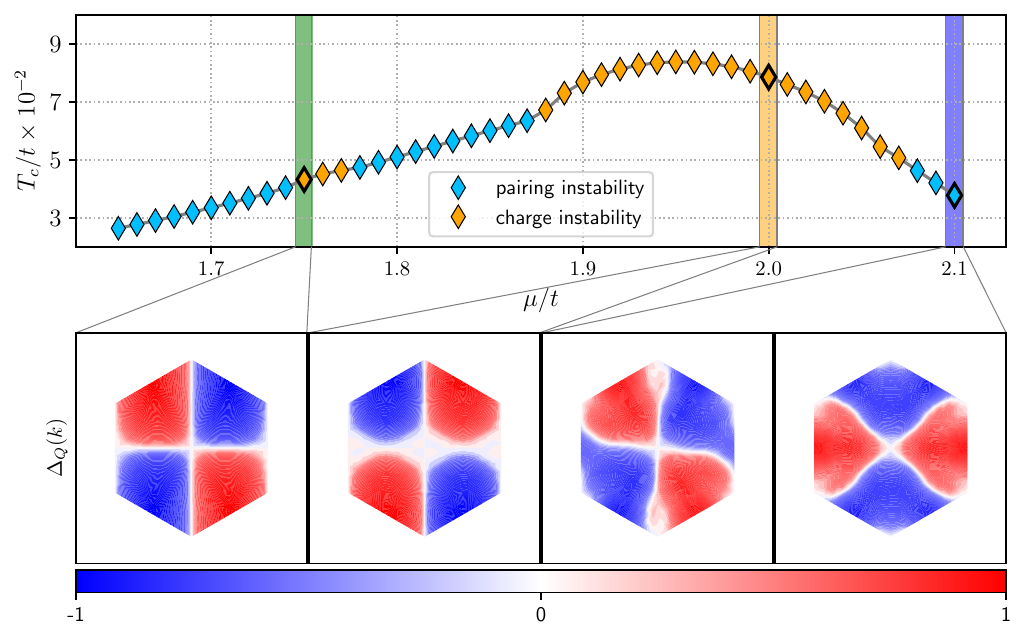}
    \caption{\textbf{FRG results for SU(4) t-U-J model} on the triangular lattice. Top: critical temperature $T_c$ of the instability as a function of $\mu$ for $U/t=3.0$ and $J/t=0.4$.
    There are three different instabilities towards charge (orange) and pairing order (blue). The charge order corresponds to $d$-wave CDWs with wave vector $M$ around $\mu=2t$ and wave vector $K/4$ around $\mu=1.75t$.
    Bottom: symmetry of the uCDW (left and middle) and pairing eigenfunctions (right) extracted from the singular FRG vertex.}
    \label{fig:frg}
\end{figure}

We show the results of the FRG analysis for $U=3t$ and $J=0.4t$ in Fig.~\ref{fig:frg}. 
We note in passing that we do not obtain a uCDW instability for $N_f=4$ and $J=0$ on the triangular lattice. 
For non-zero $J$ and varying chemical potential, we find three different instabilities. 
At and around the immediate vicinity of Van Hove filling $\mu_{\mathrm{VH}}=2t$, the leading instability occurs in the ph-d channel with wave vector $M_i$ signaling a CDW formation. To determine the form factor, we diagonalize the effective ph-d vertex $V_{\mathrm{ph-d}}(k,k',M_i)$ (see Fig.~\ref{fig:frg}), which yields the same $d$-wave form factors that rotate with the $M_i$ as described above. When the chemical potential deviates further from $\mu_{\mathrm{VH}}$, we find a transition from the $d$-wave uCDW towards a pairing instability. Diagonalization of the FRG pairing vertex gives degenerate pairing solutions with $d_{xy}$ and $d_{x^2-y^2}$ symmetry. For $1.75\leq\mu\leq 1.77$, the pairing instability is interrupted, and instead we obtain another instability in the ph-d channel, but with wave vector $K/4$. The form factors we extract from $V_{\mathrm{ph-d}}(k,k',K/4)$ in this case (see Fig.~\ref{fig:frg}) also have a $d$-wave form that matches with the RPA analysis and rotate with the six different choices for $K/4$. For $\mu>\mu_{\mathrm{VH}}$, pairing always wins over a putative $K/4$ uCDW. 

%-------
\section{Conclusion} \label{sec:conclusion}
%-------

In summary, we have put forward a mechanism for the formation of interaction-induced unconventional CDWs in metals, resulting from the condensation of particle-hole pairs with  non-zero relative angular momentum.
The generic electron-based mechanism works for repulsive Hubbard and (screened) Coulomb interactions and is reminiscent of the Kohn-Luttinger mechanism for superconductivity. 
More specifically, it takes into account one-loop pp and ph-c corrections in the bare interaction for the direct ph channel. 
This can generate attractive components in the effective interaction of the ph-d channel, leading to a charge-ordering instability with  finite wave vector when (approximate) nesting is operative.

Interestingly, this mechanism typically occurs in a regime with a strong competition of various instabilities, i.e., SDW, CDW, and pairing.
Therefore, it is generally quite challenging to identify the leading ordering tendency and, in principle, all these channels need to be taken into account on equal footing, using an unbiased many-body approach, e.g., FRG or PRG.
However, the dominance of the Kohn-Luttinger-like uCDW formation can be controlled by going to a large number of fermion flavors, which can be realized in materials platforms with, e.g., spin, orbital, and/or valley degrees of freedom.

For an application of the Kohn-Luttinger-like mechanism, we discuss the examples of square- and triangular-lattice Hubbard models at (approximate) nesting.
We numerically evaluate the effective interaction in an expansion of lattice harmonics and show the emergence of attractive uCDW components for both, the square and the triangular lattice. 
Our analysis suggests the formation of different types of $p$-wave charge density waves, classifying as loop-current or bond-ordered states.

Further, we phenomenologically explore effects beyond weak coupling at and away from Van Hove filling on both lattices by adding exchange interactions. We find $d$-wave charge density waves with  wave vector corresponding to the nesting vector as leading instabilities at Van Hove filling. 
On the square lattice, these correspond to the staggered flux state and on the triangular lattice they are loop current states, which can form an interaction-induced topological CDW phase. 
We also argue that phase transitions towards other CDWs with wave vector K/4 occur on the triangular lattice on both sides of Van Hove filling. They are towards another $d$-wave CDW for smaller $\mu$ and towards an $f$-wave CDW for larger $\mu$.
We employ an FRG approach as an independent method to confirm this result for $N_f=4$, demonstrating that the $K/4$ CDW can win the competition of orders against pairing. 

In future studies, it will be interesting to extend our formalism to multi-flavour systems relaxing SU($N_f$) symmetry for broader applications. 
We expect that charge or spin density waves with wave vector $K/4$ can be possible on other hexagonal lattices. 
Since the effect is based on the behavior of the electronic polarization, an ordinary charge density wave with wave vector $K/4$ can also occur if the coupling in the $s$-wave charge channel becomes attractive.  

\section*{Acknowledgments}
We thank Andrey Chubukov, Lorenzo Del Re, Nico Gneist, and Nikolaos Parthenios for helpful discussions. MMS acknowledges funding from the Deutsche Forschungsgemeinschaft (DFG, German Research Foundation) within Project-ID 277146847, SFB 1238 (project C02), and the DFG Heisenberg programme (Project-ID 452976698).

%-------
\begin{appendix}
%-------

%-------
\section{Form factors} \label{app:ff}
%-------

For concreteness, we list the form factors we used explicitly.
For the square lattice, they are
\begin{itemize}
    \item The on-site form factor $f_s(\mathbf{k})=  1$
    \item The nearest neighbor components: 
    \begin{align}
        f_{s'}(\mathbf{k})&=  \cos(k_x)+\cos(k_y) \notag\\
        f_{p_x}(\mathbf{k})&=   \sqrt{2}\sin(k_x) \notag\\
        f_{p_y}(\mathbf{k})&=    \sqrt{2}\sin(k_y) \notag\\
        f_{d_{x^2-y^2}}(\mathbf{k})&= \cos(k_x)-\cos(k_y)\notag
    \end{align}
    \item The next-to-nearest neighbor components: 
    \begin{align}
        f_{2s'}(\mathbf{k})&=  \cos(k_x+k_y)+\cos(k_x-k_y)\notag\\
        f_{2p_{x+y}}(\mathbf{k})&= \sqrt{2}\sin(k_x+k_y) \notag\\
        f_{2p_{x-y}}(\mathbf{k})&=  \sqrt{2}\sin(k_x-k_y) \notag\\
        f_{2d_{xy}}(\mathbf{k})&=\cos(k_x+k_y)-\cos(k_x-k_y)\notag
    \end{align}   
\end{itemize}
And for the triangular lattice
\begin{itemize}
    \item The on-site form factor $f_s(\mathbf{k})=  1$
    \item The nearest-neighbor components: 
    \begin{align}
        f_{s'}(\mathbf{k})&= \sqrt{\frac{2}{3}}(\cos(k_y)+2\cos(\frac{k_y}{2})\cos(\frac{\sqrt{3}k_x}{2})) \notag\\
        f_{f_1}(\mathbf{k})&=\sqrt{\frac{8}{3}}(\cos(\frac{k_y}{2})-\cos(\frac{\sqrt{3}k_x}{2}))\sin(\frac{k_y}{2}) \notag\\
        f_{p_y}(\mathbf{k})&=\sqrt{\frac{4}{3}}(2\cos(\frac{k_y}{2})+\cos(\frac{\sqrt{3}k_x}{2}))\sin(\frac{k_y}{2}) \notag\\
        f_{p_x}(\mathbf{k})&=2\cos(\frac{k_y}{2})\sin(\frac{\sqrt{3}k_x}{2}) \notag\\
        f_{d_{x^2-y^2}}(\mathbf{k}) &= \sqrt{\frac{4}{3}}(\cos(k_y)-\cos(\frac{k_y}{2})\cos(\frac{\sqrt{3}k_x}{2})) \notag\\
        f_{d_{xy}}(\mathbf{k})&=2\sin(\frac{k_y}{2})\sin(\frac{\sqrt{3}k_x}{2})\notag
    \end{align}
    \item The next-to-nearest-neighbor components: 
    \begin{align}
        f_{2s'}(\mathbf{k})&=\sqrt{\frac{2}{3}}(2\cos(\frac{3k_y}{2})\cos(\frac{\sqrt{3}k_x}{2})+\cos(\sqrt{3}k_x))\notag\\
        f_{f_2}(\mathbf{k})&=\sqrt{\frac{8}{3}}(\cos(\frac{3k_y}{2})-\cos(\frac{\sqrt{3}k_x}{2})\sin(\frac{\sqrt{3}k_x}{2}))\notag\\
        f_{2p_{x+y}}(\mathbf{k})&=\frac{1}{\sqrt{3}}(\sin(\sqrt{3}k_x)+\sin(\frac{\sqrt{3}(k_y-k_x)}{2})\notag\\
        &\quad\quad+2\sin(\frac{\sqrt{3}(k_y+k_x)}{2}))\notag\\
        f_{2p_{x+y}}(\mathbf{k})&=\sqrt{\frac{8}{3}}(\cos(\frac{3k_y}{2})+\cos(\frac{\sqrt{3}k_x}{2})\sin(\frac{\sqrt{3}k_x}{2}))\notag\\
        f_{2d_{x^2-y^2}}(\mathbf{k})&=\sqrt{\frac{4}{3}}(\cos(\sqrt{3}k_x)-\cos(\frac{3k_y}{2})\cos(\frac{\sqrt{3}k_x}{2}))\notag\\
        f_{2d_{xy}}(\mathbf{k})&=2\sin(\frac{3k_y}{2})\sin(\frac{\sqrt{3}k_x}{2})
    \end{align}   
\end{itemize}
On the triangular lattice at $\mathbf{Q}^\triangle$, we need linear combinations of these form factors which can be classified according to the lower symmetry. The eigenfunctions of $\hat V(\mathbf{Q}^\triangle)$ in terms of the nearest-neighbor lattice harmonics are given by
\begin{align}
    &\tilde s = s \\
    &\tilde s_-'=\frac{1}{\sqrt{3}}(\sqrt{2}s'-d_{x^2-y^2})=\cos(\mathbf{a}_2\cdot\mathbf{k})-\cos(\mathbf{a}_3\cdot\mathbf{k})\\
    &\tilde s_+'=\frac{1}{\sqrt{3}}(s'+\sqrt{2} d_{x^2-y^2}))=\sqrt{2}\cos(\mathbf{a}_1\cdot\mathbf{k})\\
    &\tilde p_{y-}=\frac{1}{\sqrt{3}}(p_y-\sqrt{2}f)=\sin(\mathbf{a}_2\cdot\mathbf{k})-\sin(\mathbf{a}_3\cdot\mathbf{k})\\
    &\tilde p_{y+}=\frac{1}{\sqrt{3}}(\sqrt{2}p_y+f)=\sqrt{2}\sin(\mathbf{a}_1\cdot \mathbf{k})\\
    &\tilde p_x =p_x\\
    &\tilde d_{xy}=d_{xy}\,. 
\end{align}
These also describe the eigenvectors of $\hat V(\mathbf{Q}^\triangle)\hat 
\Pi(\mathbf{Q}^\triangle)$, only $s\tilde s_+'\rightarrow \tilde s_+'=(3s+\sqrt{2}s'+2 d_{x^2-y^2}))/\sqrt{15}$. 

%-------
\section{Decoupling according to lattice symmetry}\label{sec:ffmixing}
%-------

In this section we show under which conditions the corrections in both, the Kohn Luttinger mechanism for pairing Eq.~\eqref{eq:Veffllpp} and the mechanism for uCDW Eq.~\eqref{eq:Veffll}
become (block-)diagonal in the irreducible representations of the lattice symmetry group. 
We start with the corrections originating in the ph channels.
We define
\begin{align}
    {\Xi}^{l,l'}_\pm \equiv \int_{\mathbf{k},\mathbf{k}'}f_l(\mathbf{k}'){\Pi}_+(\mathbf{k}\pm\mathbf{k}')f_{l'}^\ast(\mathbf{k})\,.\label{eq:Xi}
\end{align}
Comparing the mechanisms, we find that $\tau_{\mathrm{ph-c}}|_{\mathrm{ph-d}} \sim \tau_{\mathrm{ph-d}}|_{\mathrm{pp}}\sim\Xi^{l,l'}_-$, and that $\tau_{\mathrm{ph-c}}|_{\mathrm{pp}}\sim\Xi^{l,l'}_+$ differs only by a sign between the arguments of the bubble $\hat \Pi_+$. 
We can make use of the symmetry properties of the lattice harmonics we choose as form factor basis, 
which can be distributed into the irreducible  
representations of the lattice symmetry group. 
For clarity we change the way we label the form factors $l \rightarrow (\Gamma_l, \alpha)$ where $\Gamma$ labels the irreducible representation and $\alpha$ labels the basis function within that irreducible representation, i.e., it runs from 1 to the dimension $l_n$ of $\Gamma$. 
We now perform a group operation $\mathcal{R}$ which belongs to the lattice group of our system
\begin{align}
    {\Xi}^{l,l'}_\pm(0) = \int_{\mathbf{k},\mathbf{k}'}f_{\Gamma_n, \alpha}(\mathcal{R}\mathbf{k}'){\Pi}_+(\mathcal{R}(\mathbf{k}\pm\mathbf{k}'))f_{{\Gamma_{n'}, \alpha'}}^\ast(\mathcal{R}\mathbf{k})\,. 
\end{align}
The form factor  $f_{\Gamma_n, \alpha}(\mathcal{R}\mathbf{k})$ transforms like
\begin{align}
    f_{\Gamma_n, \alpha}(\mathcal{R}\mathbf{k}) =  \sum_j g_{j\alpha}^{\Gamma_n}(\mathcal{R})f_{\Gamma_n, j}(\mathbf{k}).
\end{align}
We make use of this and since the particle-hole bubble $\Pi_+(\mathbf{k})$ is symmetric under such operations $\mathcal{R}$ we obtain
\begin{align}
     &{\Xi}^{l,l'}_\pm(0) = \int_{\mathbf{k},\mathbf{k}'}\sum_{j,j'}\notag\\
     &\ \ g_{j\alpha}^{\Gamma_n}(\mathcal{R})f_{\Gamma_n, j}(\mathbf{k}')
     {\Pi}_+(\mathbf{k}\pm\mathbf{k}')
     g_{j'\alpha'}^{\Gamma_{n'}}(\mathcal{R}^{-1})
     f^\ast_{\Gamma_{n'}, j'}(\mathbf{k})
\end{align}
Now we sum over all $N_\Gamma$ symmetry operations and use the orthonormality theorem \cite{Dresselhaus2008}
\begin{align}
    {\Xi}^{l,l'}_\pm(0) &= \int_{\mathbf{k},\mathbf{k}'}\sum_{j,j'}
    \frac{1}{N_\Gamma}\sum_\mathcal{R}
    g_{j\alpha}^{\Gamma_n}(\mathcal{R})
    g_{j'\alpha'}^{\Gamma_{n'}}(\mathcal{R}^{-1})  
    \notag \\
    &\ \ f_{\Gamma_n, j}(\mathbf{k}')
    {\Pi}_+(\mathbf{k}\pm\mathbf{k}')
     f^\ast_{\Gamma_{n'}, j'}(\mathbf{k})\notag \\
     &=\int_{\mathbf{k},\mathbf{k}'}\sum_{j,j'}
    \frac{1}{l_n} \delta_{\Gamma_n, \Gamma_{n'}}\delta_{j,j'}\delta_{\alpha,\alpha'}
    \notag \\
    &\ \ f_{\Gamma_n, j}(\mathbf{k}')
    {\Pi}_+(\mathbf{k}\pm\mathbf{k}')
     f^\ast_{\Gamma_{n'}, j'}(\mathbf{k})
\end{align}
At this expression, we see that $\hat{\Xi}^{l,l'}_\pm(0)$ is of diagonal structure.
Now we look at the remaining correction
\begin{equation}
    \tau_{\mathrm{pp}}|_{\mathrm{ph-d}}^{l,l'}(\mathbf{Q})=\int_{\mathbf{k},\mathbf{k}'}f_l(\mathbf{k})f_{l'}(\mathbf{k}')\Pi_-(\mathbf{k}-\mathbf{k}'+\mathbf{Q})
\end{equation}
with $\mathbf{Q}^\square=(\pm \pi,\pm\pi)^t$ and $\mathbf{Q}^\triangle\in\{M_1,M_2,M_3\}$.
Performing again a group operation, we obtain
\begin{align}
    \Pi_-(\mathcal{R}(\mathbf{k}-\mathbf{k}')+\mathbf{Q})&= 
    \Pi_-(\mathcal{R}(\mathbf{k}-\mathbf{k}'+\mathcal{R}^{-1}\mathbf{Q})) \notag\\
    &=\Pi_-(\mathbf{k}-\mathbf{k}'+\mathcal{R}^{-1}\mathbf{Q})
\end{align}
Here we have to distinguish the two lattices. In the square lattice, $\mathbf{Q}^\square$ is invariant under the respective operations $\mathcal{R}\in C_{4v}$. For the triangular lattice, this is not the case.
This means that we can apply the orthogonality theorem after analogous steps as before for the square lattice, i.e., $V^{0,l,l'}_{\mathrm{ph-d}}(\mathbf{Q}^\square)$ is diagonal. For the triangular lattice this does not work due to the pp fluctuations. 
In this case, we can only state that even and odd form factors have no overlap and their respective components in $V^{0,l,l'}_{\mathrm{ph-d}}(\mathbf{Q}^\triangle)$ vanishes. This argumentation holds analogously for $\hat\Pi^{l,l'}_\pm$. 

%-------
\section{Patch models}\label{app:patchmodel}
%-------

\begin{figure}
    \centering
    \includegraphics[width=0.4\textwidth]{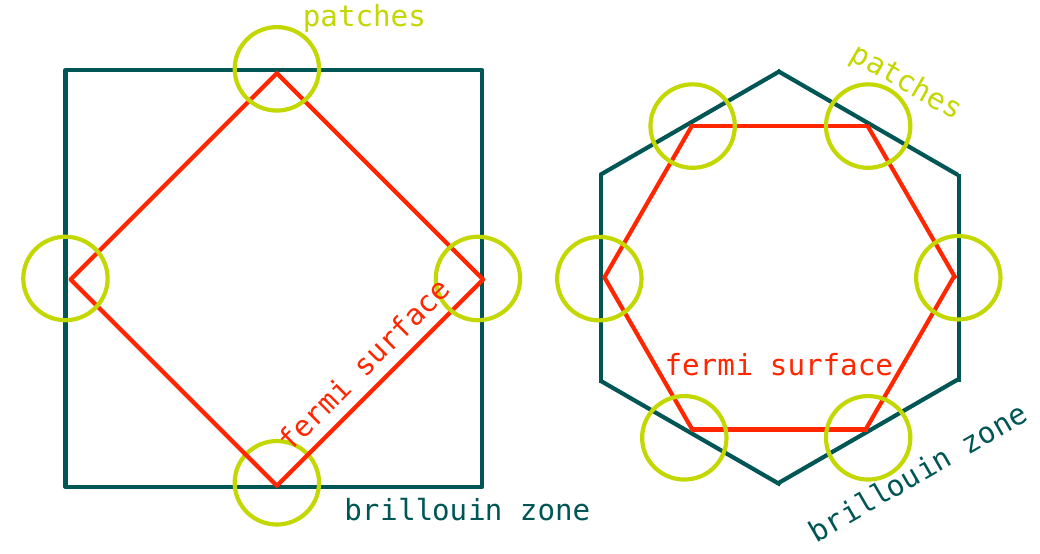}
    \caption{\textbf{Patch models} for square lattice (left) and tetragonal or hexagonal systems (right) at Van Hove filling with nested Fermi surface. Van Hove points are located inside the patches.}
    \label{fig:patches}
\end{figure}

As a general analysis, we consider effective low-energy models for tetragonal and hexagonal systems with a Van Hove singularity at the Fermi level and a nested Fermi surface. The low-energy theory is restricted to momentum states in patches around $N_p$ inequivalent Van Hove points $P_i$. For example, this situation arises on the square ($N_p=2$), triangular, honeycomb, or kagome ($N_p=3$) lattice with nearest-neighbor hopping, see Fig.~\ref{fig:patches}. The non-interacting part of the Hamiltonian is then given by
\begin{align}
    H_0=\sum_{p=1}^{N_p} \sum_{\sigma=1}^{N_f}  \sum_{\mathbf{k}}(\epsilon_{p,\mathbf{k}}-\mu)c_{\mathbf{k},p,\sigma}^\dagger c_{\mathbf{k},p,\sigma}\,,
\end{align}
with chemical potential $\mu$ and fermion annihilation (creation) operator  $c^{(\dagger)}_{\mathbf{k},p,\sigma}$ for patch $p$, flavour $\sigma$, and momentum $\mathbf{k}$. The dispersion describes saddle points $\epsilon_{1,\mathbf{k}}=a k_x^2-b k_y^2$ with $a,b>0$ and $\epsilon_{p,\mathbf{k}}$ with $p>1$ is obtained from $\epsilon_{1,\mathbf{k}}$ through rotation by $\pi/2$ for $N_p=2$ or $\pm \pi/3$ for $N_p=3$. The theory is further equipped with all symmetry-allowed, constant interactions between patches
\begin{align}
    H_V &= \frac{1}{2}\sum_{\mathbf{k}_1,\dots, \mathbf{k}_4} \sum_{\sigma, \sigma'}\delta(\mathbf{k}_1+\mathbf{k}_2-\mathbf{k}_3-\mathbf{k}_4) \notag\\ 
    &\times \left(\sum_{p\neq p'}\left[ \frac{g_1}{N_p^2} c^\dagger_{p\sigma}(\mathbf{k}_1) c^\dagger_{p' \sigma'}(\mathbf{k}_2) c_{p'\sigma'}(\mathbf{k}_3) c_{p \sigma}(\mathbf{k}_4) \right.\right.\notag \\
    &+\frac{g_2}{N_p^2} c^\dagger_{p\sigma}(\mathbf{k}_1) c^\dagger_{p' \sigma'}(\mathbf{k}_2) c_{p\sigma'}(\mathbf{k}_3) c_{p' \sigma}(\mathbf{k}_4)\notag\\
    &\left.+\frac{g_3}{N_p^2} c^\dagger_{p\sigma}(\mathbf{k}_1) c^\dagger_{p \sigma'}(\mathbf{k}_2) c_{p'\sigma'}(\mathbf{k}_3) c_{p' \sigma}(\mathbf{k}_4)\notag \right]\\
    &\left.+\frac{ g_4}{N_p} \sum_{p} c^\dagger_{p\sigma}(\mathbf{k}_1) c^\dagger_{p \sigma'}(\mathbf{k}_2) c_{p\sigma'}(\mathbf{k}_3) c_{p \sigma}(\mathbf{k}_4)\right) 
\end{align}
Low-energy effective theories based on $H_0+H_V$ were extensively studied to analyze competing orders \cite{PhysRevLett.81.3195,Nandkishore2012,HUR20091452}.  In particular, it was shown in a parquet RG analysis that there is an instability towards imaginary CDW for $N_f\geq 4$ \cite{Lin2019}. Here, we apply the KL-like analysis of the main text to the patch mode. We caution that the patch model with constant $g_i$ cannot analyze odd-$l$ harmonics because the momentum dependence around Van Hove point at $P_i$ and $-P_i$ must be distinguished for this (this is analogous to triplet pairing). 
We analyze the renormalization of the effective interaction $V_{\text{ph,d}}(k,k',Q)=V(\mathbf{k}, \mathbf{k}'+\mathbf{Q}, \mathbf{k}', \mathbf{k}+\mathbf{Q})$, which drives the CDW instability for large $N_f$. In the patch model, there are two such effective couplings $V_2=V_{\text{ph,d}}(P,P,Q)$ and $V_3=V_{\text{ph,d}}(P,P',Q)$, where $P\neq P'$ are Van Hove points and $Q=P-P'$. On the bare level, $V^0_2=g_2$ and $V^0_3=g_3$. Including pp, ph-c, and subleading ph-d corrections, we obtain 
\begin{align}
    V^0_2 &= g_2 - (2 g_2 g_4 + (N_p-2)g_2^2) \Pi_+(0) \notag \\
    &-2 g_1g_2\Pi_-(\mathbf{Q}) + 2(g_1g_2+g_3^2)\Pi_+(\mathbf{Q})\nonumber\\
     V^0_3&= g_3 + (4 g_1 g_3 +2g_2g_3)\Pi_+(\mathbf{Q}) \notag \\
     &-(2g_3g_4+(N_p-2)g_3^2)\Pi_-(0)\,. \label{eq:Vu(Q)_Vv(0)}
\end{align}
We use this as the initial interaction in the direct channel and sum the corresponding RPA series. As a result, we obtain
\begin{align}
    V_\pm=\frac{V_\pm^0}{1+N_fV_\pm^0 \Pi_+(Q)}\,,
\end{align}
where $V_\pm^{(0)}=V_2^{(0)}\pm V_3^{(0)}$. There can be an instability if $V_\pm^0<0$. For a Hubbard-like interaction with all $g_i=U$, we find $V_+^0=2U +U^2[N_p(\Pi_+(0)-\Pi_-(0))-2\Pi_-(Q)+10\Pi_+(Q)]$, and $V_-=U^2[N_p(\Pi_+(0)+\Pi_-(0))-2\Pi_-(Q)-2\Pi_+(Q)]$. For low temperatures and approximate nesting $\Pi_-(0) \gtrsim \Pi_+(Q)$, and we can neglect other contributions in comparison. In that case, $V_+^0>0$. For $V_-$, we observe a competition between ph and pp fluctuations in $V_-^0\approx U^2[N_p \Pi_-(0)- 2\Pi_+(Q)]$ and that larger numbers of patches like in the hexagonal lattices $N_p\geq 3$ prevent an attractive interaction on the one-loop level.
The coupling $V_-^0$ can at best become zero for perfect nesting $\Pi_-(0) = \Pi_+(Q)$ on the square lattice $N_p=2$. However, including higher orders of the pp and ph-c ladders would turn the sign of $V_-^0$ in favor of a CDW instability because the sign in the pp channel alternates between even and odd-loop orders. Thus, while pp and ph-c diagrams cancel for odd-loop orders they add for even-loop orders making $V_-^0<0$. 
We comment on ways to go beyond a KL approach in the main text, either via a spin fluctuation approach based on the proximity to an instability in the ph-c channel due to perfect nesting, or via the FRG which takes into account all channels on equal footing. 
We also emphasize again that we cannot make a statement on odd-$l$ harmonics within this
patch model using momentum-independent interactions $g_i$. As discussed in the main text, the KL mechanism gives an attractive coupling in the $p$-wave CDW channel on the square and triangular lattice. 

\end{appendix}
\bibliography{bibliography}
\end{document}